\documentclass[12pt]{article}
\usepackage{amsmath, amssymb}
\usepackage{graphicx}
\graphicspath{{figures/}}

\usepackage{enumerate}
\usepackage{natbib}
\usepackage{overpic}

\newcommand{\blind}{1}

\usepackage{enumitem}
\usepackage{mathrsfs, booktabs, amsthm, amsfonts, bbold}
\usepackage{hyperref}
\usepackage{cleveref}

\usepackage{accents}

\usepackage{xcolor}
\protected\def\[#1\]{\begin{equation}\begin{aligned}#1\end{aligned}\end{equation}}
\protected\def\(#1\){\begin{equation*}\begin{aligned}#1\end{aligned}\end{equation*}}

\newcommand{\exampleqed}{\hfill$\triangle$}

\newtheorem{theorem}{Theorem}

\newtheorem{lemma}{Lemma}

\theoremstyle{remark}
\newtheorem{remark}{Remark}

\theoremstyle{definition}

\newtheorem{example}{Example}
\newtheorem{condition}{Condition}[section]

\usepackage{subcaption}
\usepackage{float}

\newcommand{\AND}{\\[4pt]}

\begin{document}

\def\spacingset#1{\renewcommand{\baselinestretch}
{#1}\small\normalsize} 
\spacingset{1}

\if1\blind
{
  \title{Mirror Likelihood for Modeling Manifold Data \\with Application on Correlation Time Series
  }
  \author{
    Luis Hasenauer\thanks{\href{email:l.hasenauer@ufl.edu}{l.hasenauer@ufl.edu}}\\
    Department of Statistics, University of Florida
    \\
    \AND \\
    Arkaprava Roy\thanks{\href{email:arkaprava.roy@ufl.edu}{arkaprava.roy@ufl.edu}}\\
  Department of Biostatistics, University of Florida\\  
    \AND\\
      Leo Duan\thanks{\href{email:li.duan@ufl.edu}{li.duan@ufl.edu}}\\
    Department of Statistics, University of Florida\\  
    }
    \date{}
  \maketitle
  \vspace{-0.5in}
} \fi

\if0\blind
{
  \bigskip
  \bigskip
  \bigskip
  \begin{center}
    {\LARGE\bf }
\end{center}
  \medskip
} \fi

\bigskip

\begin{abstract}
  Manifold-valued data arise in a wide range of scientific applications. Existing statistical methods typically follow either an intrinsic approach, which characterizes the manifold itself via quantities such as geodesic distance or the Fr\'echet mean, or an extrinsic approach, which embeds the manifold into an ambient Euclidean space, performs unconstrained statistical estimation, and projects the result back to the manifold. 
  While the latter is conceptually and computationally simple, likelihood-based extrinsic solutions remain scarce, due to the lack of a general recipe for choosing an embedding map and for characterizing its impact on the transformed distribution.
  In this article, we tackle the problem using an augmented likelihood approach: we view the manifold-valued data as the Fenchel mirror map of a latent variable, which is nearly unconstrained, and model the latent variable with a canonical distribution, such as Gaussian. The key transform, the Fenchel mirror map, is a generalization of the Euclidean projection and can effectively map into the interior of the manifold. Taking a Bayesian approach, the proposed framework enjoys a generative model interpretation and an augmented likelihood free from intractable normalization constants, and it supports principled uncertainty quantification and flexible model-based extensions. We study the theoretical and computational properties of the framework and demonstrate its utility in analyzing neuroimaging data of high-dimensional correlation time series, where each correlation matrix is rank deficient.
  \end{abstract}

\section{Introduction}

Technological advances in sciences routinely produce data whose support is not a Euclidean space but a manifold.
Directional observations on circles and spheres have been studied since \cite{fisher1953dispersion} on geomagnetic pole directions and remain central to spatial statistics~\citep{mardia2000directional}.
Beyond spheres, manifold-valued data appear as shape configurations of anatomical landmarks~\citep{kendall1984shape,dryden1998statistical}, as diffusion tensors in neuroimaging~\citep{alexander2007diffusion}, as low-dimensional subspaces in signal processing and machine learning~\citep{chikuse2003statistics}, and as orthonormal frames, curves, and networks~\citep{kolaczyk2017averages}.
In all these settings, ordinary Euclidean operations are not well defined on the data space: the arithmetic mean of points on a sphere lies in the interior rather than on the surface, Euclidean distances between shape configurations confound rotation with morphological variation, and linear operations on angular quantities are discontinuous at the period boundary. Consequently, a statistical model that ignores the manifold constraint may yield estimates that lie outside the sample space, distort measures of dispersion and association, and produce predictions that violate the geometry of the data~\citep{bhattacharya2012nonparametric}.

Existing statistical methods for manifold-valued data fall broadly into two paradigms.
Intrinsic approaches work directly on the manifold using geodesic distances, Fr\'echet means, and Riemannian log maps~\citep{bhattacharya2003large,bhattacharya2005large}. These tools have been extended to nonparametric density estimation~\citep{pelletier2005kernel}, geodesic and Fr\'echet regression~\citep{fletcher2011geodesic, petersen2019frechet}, and intrinsic local polynomial models~\citep{yuan2012local}.
While theoretically elegant, intrinsic methods face practical obstacles: optimal estimators often lack closed-form expressions, the required optimization must respect the manifold constraint at every step, and the resulting objective functions are frequently non-convex, so convergence guarantees can be difficult to obtain~\citep{lin2017extrinsic}.
Extrinsic approaches circumvent these difficulties: they embed the manifold into an ambient Euclidean space, carry out standard statistical procedures in the ambient space, and project the resulting estimate back onto the manifold, thereby retaining the tractability of Euclidean inference.

In the literature on extrinsic methods, early works focused on moment-based inference: given an embedding $J: \mathcal M \to \mathbb{R}^D$, one forms the Euclidean mean of the embedded observations and projects it back onto the manifold $\mathcal M$, yielding an extrinsic sample mean~\citep{bhattacharya2003large,bhattacharya2005large}. The embedding and projection are manifold-specific. For example, for the sphere $S^d$, one embedding is simply inclusion $J(x)=x$ with radial projection $y \mapsto y/\|y\|$; for the planar shape space $\Sigma_2^k$, the Veronese--Whitney embedding maps a shape to a complex Hermitian matrix, followed by projection onto the dominant eigenvector; for the Grassmannian $\mathrm{Gr}_k(\mathbb{R}^m)$, the embedding is $J(\sigma(X)) = XX^\top$ into symmetric matrices, followed by projection by top-$k$ eigendecomposition~\citep{bhattacharya2012nonparametric}. \citet{bhattacharya2010nonparametric} proposed a Bayesian nonparametric density model on $\Sigma_2^k$ using Dirichlet process mixtures of complex Watson kernels, where the kernel distance is the chord distance induced by the Veronese--Whitney embedding. For regression with manifold-valued responses, \citet{lin2017extrinsic} developed extrinsic kernel and local polynomial estimators for several manifolds, including spheres, planar shape spaces, the Stiefel manifold, and Grassmannians. On the Stiefel manifold, \citet{lin2017bayesian} used Dirichlet process mixtures of matrix Langevin kernels under the inclusion embedding, while \citet{stthomas2014learning} employed the Grassmann embedding $XX^\top$ for subspace learning, exploiting the extrinsic Frobenius distance $\|X_1X_1^\top - X_2X_2^\top\|_F$ as a principled metric between subspaces of differing dimension. \citet{lin2019extrinsic} later constructed extrinsic Gaussian process priors by defining a covariance kernel on the image of an embedding as $K_{\mathrm{ext}}(x_1, x_2) = \tilde{K}(J(x_1), J(x_2))$ with any standard Euclidean kernel $\tilde{K}$.

We are particularly inspired by a subclass of extrinsic methods, namely, the sphere-projected Gaussian models for directional data \citep{presnell1998projected}. As the name implies, these models are obtained by radially projecting a latent Gaussian onto the sphere. Subsequent works extend to general projected Gaussian with unrestricted covariance~\citep{wang2013directional}, to spatial and space-time dependence through projected Gaussian processes~\citep{wang2014modeling}, to nonparametric Dirichlet process mixtures~\citep{nunez2015bayesian}, and to arbitrary ambient dimension~\citep{hernandez2017general}. Compared to other classical extrinsic solutions, the projected Gaussian construction has some unique advantages. More specifically, classical extrinsic methods begin with a fixed embedding of the manifold and use the associated projection only to map a single ambient estimate back to the manifold; the projected Gaussian model instead specifies a distribution on the latent ambient variable and takes the manifold-valued observation to be its image under the projection, so that the projection is part of the data-generating mechanism rather than a post-estimation step.
Since the projection from the latent space onto the manifold can be multiple-to-one, the multiplicity is correctly accounted for through the Jacobian factor, enabling coherent downstream inference. In other words, the strength of sphere-projected Gaussian comes from the likelihood framework.

We adopt this likelihood-based perspective as our starting point and extend it beyond the sphere to general manifolds. In particular, we are motivated by the task of modeling correlation tensors (sequences of correlation matrices) that arise in neuroscience studies. A substantial Bayesian literature addresses the modeling of a covariance matrix \citep{sagar2024precision,lee2026scalable,bersson2025bayesian,yang2024empirical,lewandowski2009generating,berger2020bayesian,bhadra2024evidence}. Compared with covariance, correlation matrix data analysis usually faces many complications at the same time, including unit diagonal constraints, rank deficiency, rank variation, and sparsity; hence new work is needed.

There are three challenges in our effort of generalization. First, beyond the sphere there is in general no canonical choice of embedding and projection, so the construction lacks a default solution. Second, the Euclidean projection can be poorly suited to the geometry of the target space, since it cannot connect an external point to an interior point of the manifold. For example, a strictly positive definite (SPD) matrix resides in the interior of the positive semi-definite cone, hence the only possible projection of such a matrix is the matrix itself, rendering the projection mapping useless. Third, for likelihood-based inference the preimage of a point under the projection may be analytically intractable, making the manifold distribution difficult to characterize. We address these three challenges using a tool from the optimization literature, the Fenchel mirror map.

Generalizing the canonical Euclidean projection, the Fenchel mirror map employs a generator function to effectively connect the interior of the manifold and the Euclidean space: the generator function can create a repulsion (if needed) from the boundary, ensuring that the transformed point always lies in the relative interior. Taking a Bayesian approach, we model a latent Euclidean variable with a canonical likelihood and treat the observed manifold point as its Fenchel mirror map, yielding an augmented likelihood that is free of intractable normalization constants. The framework admits a natural generative interpretation: one draws a latent vector, then applies the Fenchel mirror map to obtain a point on the manifold. As the word {\em mirror} suggests, one could view the Euclidean latent variable as a mirror image of the manifold-valued observation. We show that this intuitive modeling framework supports principled uncertainty quantification through posterior credible regions and enables flexible model-based extensions.

\section{Methodology}
\subsection{Mirror likelihood}
Suppose our data of interest $y_i\in \mathcal M\subseteq \mathbb R^d$ are points on a manifold $\mathcal M$. By a manifold, we mean a set that locally resembles a Euclidean space near each of its points. We use the term in a broad sense: $\mathcal M$ may be a union of sets of varying dimension, and it need not be smooth or convex. To properly motivate the Fenchel mirror map, we first focus on the case when $y_i$ resides in the {\em interior}, or {\em relative interior}, of $\mathcal M$: 
\(
& \operatorname{int}(\mathcal M) = \{y \in \mathcal M : \exists\, \varepsilon > 0 \text{ such that } B(y,\varepsilon) \subseteq \mathcal M\},\\
&\operatorname{ri}(\mathcal M) = \{y \in \mathcal M : \exists\, \varepsilon>0 \text{ such that } B(y,\varepsilon)\cap \operatorname{aff}(\mathcal M) \subseteq \mathcal M\},
\)
with $\operatorname{aff}(\mathcal M)$ the affine hull of $\mathcal M$, and $B$ an open ball. The relative interior is a useful concept generalizing the interior, $\operatorname{int}(\mathcal M)\subseteq\operatorname{ri}(\mathcal M)$, since a set such as the probability simplex $\Delta^{d-1}=\{w\in\mathbb{R}_{\ge 0}^d: \sum_{j=1}^d w_j=1 \}$ can have an empty $\operatorname{int}(\mathcal M)$ but a non-empty $\operatorname{ri}(\mathcal M)$. Further, we use $\operatorname{bd}(\mathcal M)$ to denote the boundary, which is the closure of $\mathcal M$ minus $ \operatorname{int}(\mathcal M)$. Modeling $y_i$ in $\mathcal{M}$ is challenging, as there is a scarcity of applicable and tractable models for such a space.

\begin{figure}[H]
    \centering
    \begin{subfigure}[t]{0.45\textwidth}
        \centering
        \includegraphics[width=0.7\textwidth]{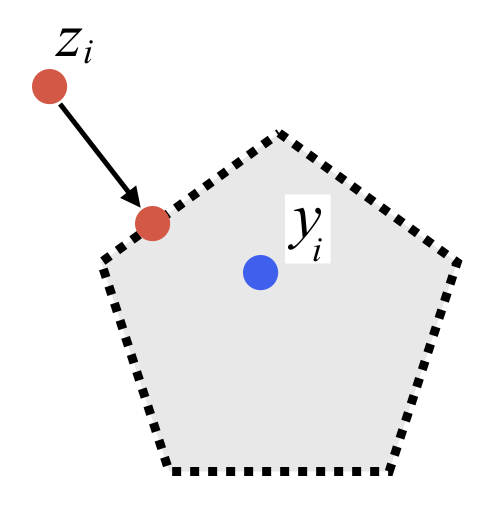}
        \caption{Euclidean projection: for some $\mathcal M$, the less restricted $z_i$ outside $\mathcal M$ are projected to the boundary of $\mathcal M$, hence $y_i$ in the interior of $\mathcal M$ cannot be viewed as the output of a projection. }
    \end{subfigure}
    \qquad
    \begin{subfigure}[t]{0.45\textwidth}
        \centering
        \includegraphics[width=0.7\textwidth]{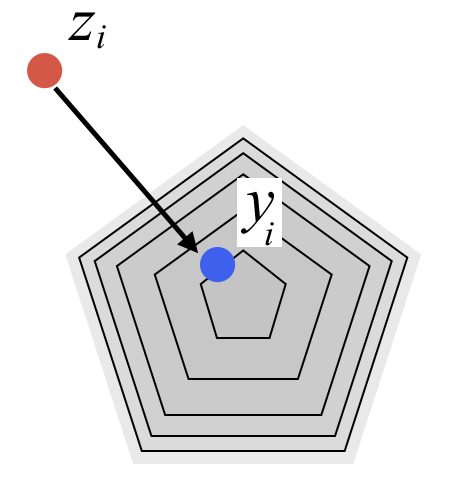}
        \caption{Fenchel mirror map: a generator function $\psi$ can create a continuous repulsion (contour lines) from the boundary, pushing the output into the relative interior of $\mathcal M$, connecting $y_i\in \operatorname{ri}(\mathcal M)$ and $z_i\in \mathbb R^d$.}
        \label{fig:fenchel_mapping}
    \end{subfigure}
    \caption{
        Illustration of Euclidean projection and its generalization to Fenchel mirror map for relaxing constraints in modeling data on a manifold $\mathcal{M}$ (gray region).
    }
    \label{fig:fenchel_vs_projection}
\end{figure}

It is natural to consider a canonical projection approach, by modeling $y_i$ as the Euclidean projection of an unrestricted latent $z_i\in \mathbb R^d$ onto $\mathcal M$: $y_i = \arg\min_{w\in \mathcal M} \|w - z_i\|_2^2$. Nevertheless, it is not hard to see that when $z_i\not\in \mathcal M$, the minimizer can only be in $\operatorname{bd}(\mathcal M)$. As a result, for $y_i\in \operatorname{int}(\mathcal M)$, the only possible $z_i$ may be $y_i$ itself, which is not useful for the purpose of relaxing constraints. See Figure \ref{fig:fenchel_vs_projection}(a).

This motivates us to consider a more general data generative mechanism:
\[\label{eq:fenchel_conjugate}
& z_i \stackrel{\text{indep}}{\sim} \phi(\cdot\,;\theta_i),\\
& y_i = T(z_i)=\arg\max_{w\in\mathbb{R}^d}\; \bigl\{\langle z_i, w\rangle - \psi(w)\bigr\}.
\]
 We use $\phi(\cdot\,;\theta_i)$ to denote both the distribution and the density of a latent variable $z_i\in \mathbb R^d$, with $\theta_i$ its parameter that could vary with covariates. Despite conditional independence, the above allows marginal dependence among $(z_1,\dots,z_n)$ through $\theta_i$.

The function $\psi:\mathbb{R}^d\to\mathbb{R}\cup\{+\infty\}$ is a closed and proper function that we call the generator function. We use the following form:
\[\label{eq:generator_form}
\psi(w) = h(w) + \iota_{\mathcal M}(w),
\]
where $h:C\to\mathbb{R}$ is a finite-valued function with domain $C\supseteq \mathcal M$, and $\iota_{\mathcal M}(w)$ is the set indicator function, which takes value 0 if $w\in \mathcal M$ and $\infty$ otherwise. This function is chosen according to the geometry of the manifold so that $\psi(w)$ has appropriate behavior for $w\in \mathcal M$ near the boundary. For example, for strictly positive $y_i$, $h(w)= -\sum_{j=1}^d \log(w_j)$ tends to $\infty$ as $w_j\downarrow 0$, hence preventing a zero-valued output coordinate when such a value is impossible in the observation space. On the other hand, $h(w) = \frac12\|w\|_2^2$ yields $T$ as the Euclidean projection; therefore, \eqref{eq:fenchel_conjugate} is a generalization of the projection-based model.

The transformation $T$ is the Fenchel mirror map, related to the Fenchel conjugate of $\psi$:
\[\label{eq:fenchel_conjugate_def}
    \psi^*(z_i) = \sup_{w\in\mathbb{R}^d}\; \bigl\{\langle z_i, w\rangle - \psi(w)\bigr\}.
\]
That is, when $\psi^*(z_i)$ is finite and attainable by some finite $w$, $T(z_i)$ is the maximizer. We focus on the case when the maximizer in \eqref{eq:fenchel_conjugate} is unique, except for a set of $z_i$ with zero Lebesgue measure under the density $\phi(\cdot\,;\theta_i)$.

\begin{remark}
    In a rather restricted setting with $\psi$ being strictly convex and essentially smooth (i.e., Legendre), we have $y_i = T(z_i) = \nabla\psi^*(z_i)$ and $z_i= T^{-1}(y_i)= \nabla\psi(y_i)$; hence $y_i$ and $z_i$ are, in effect, mirror images of each other, and we name $T$ the forward Fenchel mirror map (mirror map for short), although we focus on much more general cases when $\psi$ is not necessarily Legendre, and $T$ is not necessarily one-to-one.
\end{remark}

We have an augmented likelihood called the {\em mirror likelihood}:
\[\label{eq:mirror-likelihood}
    \mathcal L(y_1,\dots,y_n, z_1,\dots,z_n; \theta) = \prod_{i=1}^n \phi(z_i;\theta_i)\, \mathbf 1\big\{ y_i =T(z_i) \big\},
\]
where $\mathbf 1(\cdot)$ is the $0$--$1$ indicator function. It is worth noting that the augmented likelihood does not contain additional normalizing constants, which is a strength of the proposed framework. Marginally, we have
\[
    \mathcal L(y_i; \theta) = \int_{\mathcal F_{y_i}} \phi(z;\theta_i) \, \nu(\text{d}z),
\]
where $\nu$ is the appropriate measure on the fiber $\mathcal F_{y_i}=\{z\in \mathbb R^d: T(z)=y_i\}$, which can correspond to a Lebesgue measure, a counting measure, or a point mass. For example, when $\psi$ is Legendre, the mapping is one-to-one, hence $\nu$ is a point mass. For general cases, the marginal likelihood can be intractable, but we can take a Bayesian approach and use the data augmentation technique \citep{tanner1987calculation} to handle the computation.

\begin{remark}
    \label{rem:jacobian}
To prevent confusion on the Jacobian term, when we write $\mathcal L(y_i,z_i;\theta)$, the only continuous density involved is that of the latent $z_i$; the observation $y_i$ is the deterministic image $T(z_i)$ and so contributes a point mass. Its conditional law given $z_i$ is a Dirac measure, which carries no density and hence no Jacobian. Second, the posterior computation of Section~\ref{sec:computation} does not sample $z_i$ in the ambient coordinates but in a coordinate of the fiber $\mathcal F_{y_i}$, writing $z_i = g(\eta_i)$ for an unconstrained $\eta_i$. This is a change of variables on the latent space, so the target for $\eta_i$ must include the Jacobian of $g$.
\end{remark}

For illustration, we now use a simple manifold $\mathcal M = \Delta^{d-1} = \{w\in\mathbb R_{\ge 0}^d: \sum_{j=1}^d w_j=1 \}$, the probability simplex. Suppose our interest is in the relative interior of the simplex where all elements are strictly positive.

\begin{example}[Probability simplex]
    \label{ex:entropy-simplex}
We use Shannon entropy $\psi(w) = \sum_{j=1}^d w_j\log w_j + \iota_{\Delta^{d-1}}(w)$, with convention $0\log 0 = 0$. The Fenchel conjugate is the log-sum-exp function $\psi^*(z_i) = \log\bigl(\sum_{j=1}^d e^{z_{i,j}}\bigr)$ defined on $\mathbb R^d$, and the mirror map is the softmax transformation
    \(
        y_{i,j} = {e^{z_{i,j}}}/(\sum_{k=1}^d e^{z_{i,k}}), \qquad j=1,\dots,d.
    \)
    \exampleqed
    \end{example}    

From the last example, we note two properties that are general to many other manifolds. First, the generator $\psi(w)$ does not always have to diverge to prevent $y_{i,j}$ from landing on the boundary or relative boundary (e.g., $y_{i,j}=0$, as $0\log 0 = 0$ is non-diverging). To understand how this works, we note that the first-order optimality condition for the maximizer $w^*\in \mathcal M$ is given by
\(
    \langle z_i - \nabla \psi(w^*) , v-w^* \rangle\le 0 \;\;\forall v\in \mathcal M.
\)
In the example, $-(\nabla \psi(w))_j = -1- \log w_j \to \infty$ as $w_j\downarrow 0$, and $v_j-0>0$ would prevent zero-valued $w_j$ from being optimal. Intuitively, the derivative $-\nabla \psi(w)$ is a diverging repulsion force that pushes $w_j$ away from zero.

Second, the transformation of $z_i$ to $y_i$ does not have to be one-to-one. In the example, since $\operatorname{int}(\Delta^{d-1}) = \varnothing$, $\psi(w)$ is not Legendre, and the fiber is not a singleton: if $z_i \in \mathcal F_{y_i}$, $z_i + c\mathbf 1_d$ with $c \in \mathbb R$ is also in $\mathcal F_{y_i}$.
    
 With a suitable prior, the posterior can be estimated efficiently using data augmentation Markov chain Monte Carlo (DA-MCMC) algorithms, which alternately samples $z_i$ and $\theta_i$. On the other hand, the fiber $\mathcal F_{y_i}$ may have a complicated geometry, which makes the sampling of $z$ {\em appear} difficult. This motivates some more careful characterization of the fiber, using the optimality conditions of the Fenchel conjugate.

\subsection{Fiber of the mirror map}
\label{subsec:fiber}
As mentioned above, even though $T$ is surjective almost everywhere, it is not necessarily one-to-one. Fortunately, we still have a simple characterization of the fiber, $\mathcal F_{y_i}$, regardless of whether $\psi$ and $\mathcal M$ possess nice properties such as convexity or smoothness.
\begin{theorem}
    \label{thm:subdifferential-guarantee}
For any function $\psi: \mathbb R^d \to \mathbb R \cup \{+\infty\}$, define the subdifferential at $y_i$ as $
    \partial \psi(y_i) = \{g: \psi(w) \ge \psi(y_i) + \langle g, w-y_i \rangle \;\;\forall w \in \mathbb R^d\}
$, then 
\[\label{eq:subdifferential-guarantee}
y_i = \arg\max_{w\in \mathbb R^d} \left\{ \langle z_i, w\rangle - \psi(w) \right\}\; \Leftrightarrow \; z_i \in \partial \psi(y_i).
\]
\end{theorem}
The above equivalence allows us to use subdifferential as a mathematical tool to analyze the fiber. Next, one may question if the fiber $\mathcal F_{y_i}$ is empty, especially when the manifold $\mathcal M$ is non-convex. The following dispels this concern, provided we have a suitable $h$.

\begin{theorem}
    \label{thm:fiber-nonempty}
For $\psi(w)=h(w) + \iota_{\mathcal M}(w)$ and $y_i=\arg\max_{w} \left\{ \langle z_i, w\rangle - \psi(w) \right\}$, if $h$ is convex and finite over $C$, and $\mathcal M\subseteq \text{ri}(C)$, then $\mathcal F_{y_i}$ is non-empty for any $y_i\in\mathcal M$.
\end{theorem}
Note that $h$ is convex, but $\mathcal M$ does not need to be. The non-empty fiber for any $\mathcal M$ is especially encouraging: $\mathcal M$ can be non-continuous or non-smooth sets. This is in contrast to traditional manifold models, such as tangent-space models, whose Riemannian exponential map is limited to smooth manifolds.

Further, when $\psi$ is the sum of two convex functions, its subdifferential is separable, giving a convenient way to find the fiber.
To be concrete, consider $\psi(w) = h(w) + \iota_{\mathcal M}(w)$, with $h$ being convex and differentiable over $C\supseteq \mathcal M$, and $\mathcal M$ being a closed convex set:
\[\label{eq:convex-fiber}
    \partial \psi(y_i) = \nabla h(y_i) + N_{\mathcal M}(y_i),
\]
where $N_{\mathcal M}(y_i):=\{g: \langle g, w-y_i \rangle \le 0 \;\;\forall w \in \mathcal M\}$ is the normal cone of $\mathcal M$ at $y_i$ --- that is, the set of vectors making a non-acute angle with every direction into $\mathcal M$ from $y_i$. We now illustrate this result using the probability simplex example.

\addtocounter{example}{-1} 
\begin{example}[Probability simplex, continued]
Using Example \ref{ex:entropy-simplex}, for $\mathcal M=\Delta^{d-1}$, we have 
$$N_{\Delta^{d-1}}(y_i) = \{c_i\mathbf 1_d: c_i \in \mathbb R\}.$$ If all $y_{i,j}>0$, we recover the fiber $\mathcal F_{y_i} = \{  (\log y_{i,j} +1 + c_i)_{j=1}^d: c_i \in \mathbb R \}$. Suppose we assign independent log-gamma densities to each coordinate of $z_i$, with $f(z_{i,j}; \alpha_j, \beta_j) \propto \exp(\alpha_j z_{i,j} - \beta_j e^{z_{i,j}})$ for shape $\alpha_j>0$ and rate $\beta_j>0$. Using the fiber constraint $z_{i,j} = \log y_{i,j} + 1 + c_i$, the mirror likelihood for each $y_i$ is
\(
    \mathcal{L}(y_i, z_i; \theta) \propto \Bigl(\prod_{j=1}^d y_{i,j}^{\alpha_j}\Bigr) \exp\!\Bigl(d\bar\alpha (1+c_i) - e^{1+c_i}\sum_{j=1}^d \beta_j y_{i,j}\Bigr) \, \mathbf 1(z_i\in\mathcal F_{y_i}),
\)
where $\bar\alpha = \frac1d\sum_{j=1}^d\alpha_j$. To clarify, although the right hand side above uses $c_i$, it is still the joint density of $(y_i, z_i)$, with $z_i$ evaluated via the fiber parameterization. Transforming into the joint of $(y_i,c_i)$ and integrating out $c_i$, we have
\(
\mathcal{L}(y_i; \theta) \propto \int_{\mathbb R} \Bigl(\prod_{j=1}^d y_{i,j}^{\alpha_j-1}\Bigr) \exp\!\Bigl(d\bar\alpha (1+c_i) - e^{1+c_i}\sum_{j=1}^d \beta_j y_{i,j}\Bigr) dc_i\propto \Bigl(\prod_{j=1}^d y_{i,j}^{\alpha_j-1}\Bigr) \Bigl(\sum_{j=1}^d \beta_j y_{i,j}\Bigr)^{-d\bar\alpha},
\)
which generalizes the Dirichlet distribution, recovered when all $\beta_j\equiv \beta_0>0$.
\exampleqed
\end{example}

It is natural to expect complications for general cases with non-convex $h$ or $\mathcal M$. Indeed, there may no longer be an analytical solution for $\mathcal F_{y_i}$; however, the following result shows the convexity of $\mathcal F_{y_i}$.
\begin{theorem}
    \label{thm:fiber-nonconvex}
For any given $y_i\in\mathcal M$, the fiber
$
    \mathcal F_{y_i} = \bigl\{z\in\mathbb R^d:\ \langle z,\,w-y_i\rangle \;\le\; \psi(w)-\psi(y_i) \; \forall w\in\mathcal M\bigr\}$ is convex. Further,
$\mathcal F_{y_i}$ is the set of minimizers of
$
    F_{y_i}(z) = \psi^*(z) - \langle z, y_i\rangle.
$
\end{theorem}
The convexity of $\mathcal F_{y_i}$ immediately enables tractable sampling of $z_i$ in the posterior computation.
There are many efficient algorithms for drawing Markov chain samples in a convex set, such as hit-and-run \citep{smith1984efficient,belisle1993hit,lovasz1999hit} or ball walk algorithms \citep{kannan2009random}.

\subsection{Models for common non-convex manifolds}
\label{sec:models-nonconvex}
We present a few examples of mirror likelihoods for non-convex $\mathcal M$, which are common in the statistical literature. The Stiefel manifold of orthonormal frames arises whenever the object of interest is an orthogonal factor: principal component loadings, factor and reduced-rank regression models, and the Procrustes alignment in shape analysis \citep{dryden1998statistical}. Quotienting out the rotation within a frame gives the Grassmann manifold of subspaces, the natural parameter space when only the span is identifiable, as in sufficient dimension reduction and subspace modeling  \citep{chikuse2003statistics}.

\begin{example}[Stiefel manifold]
    \label{ex:stiefel}
Let $\mathcal M = \{Y \in \mathbb R^{d\times k}: Y^\top Y = I_k\}$ be the Stiefel manifold of orthonormal $k$-frames in $\mathbb R^d$, with $k\le d$. With $\psi = \iota_{\mathcal M}$, the mirror map is
$
    T(Z) = \arg\max_{W\in\mathcal M}\operatorname{tr}(Z^\top W).
$
This is the classical orthogonal Procrustes problem, whose maximum is attained at $Y= W = U_Z V_Z^\top$ where $Z = U_Z\Sigma_Z V_Z^\top$ is a compact SVD; equivalently, $Y$ is the orthogonal factor in the polar decomposition
\(
    Z = \underbrace{U_ZV_Z^\top}_{Y}\,\underbrace{V_Z\Sigma_ZV_Z^\top}_{P}.
\)
The maximizer is unique when $Z$ is full rank, equivalent to $P$ being positive definite. Therefore, we have the fiber 
\(
    \mathcal F_Y = \bigl\{ Y P: P\succ 0\bigr\},
\)
which is convex even though $\mathcal M$ is non-convex.

With a matrix-normal density $Z_i\sim\mathcal N(M,\sigma^2 I, I)$ on $\mathbb R^{d\times k}$,  the mirror likelihood is 
\(
    \mathcal L(Y_i, Z_i; M,\sigma^2) \propto {\sigma}^{-dk} \exp\!\bigl(-\tfrac{1}{2\sigma^2}\|Y_iP_i - M\|_F^2\bigr).
\)
Again, the right hand side is the joint density of $(Y_i, Z_i)$, with $Z_i$ evaluated via the fiber parameterization (similar rule applies in all the following examples, hence we omit the explanation from now on). From the above we can see that,
 (i) the conditional distribution $\mathcal L(Y_i \mid P_i, M, \sigma^2)$ is a matrix-von Mises–Fisher  $vMF(MP_i^\top /\sigma^2)$ \citep{khatri1977mises}, whose normalizing constant is intractable due to the hypergeometric function of matrix argument \citep{lin2017bayesian}; 
 (ii) the normalizing constant is canceled out in the joint likelihood $\mathcal L(Y_i, Z_i; M,\sigma^2)$, making the posterior computation tractable. A similar transformation was previously used by \citet{jauch2021monte} for sampling on the Stiefel manifold; our goal here is different in that we observe the manifold-valued $Y_i$ and aim to find $Z_i: T(Z_i) = Y_i$.
\exampleqed
\end{example}

\begin{example}[Grassmann manifold]
    \label{ex:grassmann}
Let $\mathcal M = \{Y \in \mathbb S^d:\ Y^2 = Y,\ \operatorname{tr}(Y) = k\}$ be the set of rank-$k$ orthogonal projections, giving the Grassmann manifold of $k$-dimensional subspaces of $\mathbb R^d$. With $\psi = \iota_{\mathcal M}$ and $W = \sum_{j=1}^k u_j u_j^\top$ with $\{u_j\}$ orthonormal, the mirror map is
\(
    T(Z) = \arg\max_{W\in\mathcal M}\operatorname{tr}(ZW)= \arg\max_{W\in\mathcal M} \sum_{j=1}^k u_j^\top Z u_j.
\)
The maximum is attained when $\{u_j\}$ spans a top-$k$ eigenspace of $Z$, giving $\sum_{j=1}^k\lambda_{j}(Z)$, with $\lambda_j$ the eigenvalues in non-increasing order. The maximizer is unique when $\lambda_k(Z) > \lambda_{k+1}(Z)$.

Writing $Y = UU^\top$ with $U = [u_1,\dots,u_k]$ satisfying $U^\top U = I_k$, and $U_\perp\in\mathbb R^{d\times(d-k)}$ an orthonormal basis of $\operatorname{range}(U)^\perp$, we have
\(
    \mathcal F_Y = \bigl\{ U A U^\top + U_\perp B U_\perp^\top:\
    A\in\mathbb S^{k},\ B\in\mathbb S^{d-k},\ \lambda_{\min}(A)\ge\lambda_{\max}(B)\bigr\} .
\)

Here $Z_i$ is symmetric. We take the density $\propto\exp\bigl(-\|Z_i - M\|_F^2/(2\sigma^2)\bigr)$ with respect to Lebesgue measure on the free coordinates $\{Z_{i,jl}\}_{j\le l}$, that is, $Z_{i,jj}\sim\mathcal N(M_{jj},\sigma^2)$ and $Z_{i,jl}\sim\mathcal N(M_{jl},\sigma^2/2)$ independently for $j<l$, with $M\in\mathbb S^d$. The mirror likelihood is
\(
    \mathcal L(Y_i, Z_i; M,\sigma^2) \propto \sigma^{-d(d+1)/2}\exp\!\bigl(-\tfrac{1}{2\sigma^2}\|U_i A_i U_i^\top + U_{\perp,i} B_i U_{\perp,i}^\top - M\|_F^2\bigr).
\)
Writing $O_i = [U_i, U_{\perp,i}]\in O(d)$ and $\Lambda_i = \operatorname{diag}(A_i, B_i)$ and expanding the square leaves
$    \mathcal L(O_i \mid A_i, B_i, M, \sigma^2) \propto \exp\bigl\{\tfrac{1}{\sigma^2}\operatorname{tr}\bigl(\Lambda_i\, O_i^\top M O_i\bigr)\bigr\},$
a matrix Bingham distribution \citep{jupp1979maximum,hoff2009simulation}, from which $Y_i = U_iU_i^\top$ is recovered deterministically. The normalizing constant of the matrix Bingham distribution is intractable but cancels out in the joint likelihood, as in the Stiefel manifold example. Note that $(M,\sigma^2)$ and $(aM,\,a^2\sigma^2)$, for any $a>0$, induce the same law for $Y_i$; therefore, we hold $\sigma^2=1$.
\exampleqed
\end{example}

In some cases, we have an expression for the fiber, but drawing samples is not straightforward (such as in the above example, $\lambda_{\min}(A)\ge\lambda_{\max}(B)$ is difficult to satisfy via reparameterization). We will handle this issue using algorithms described in the computation section.

\subsection{Center and dispersion}
\label{sec:center}

The mirror map $T$ pushes the continuous distribution of the latent variable forward onto $\mathcal M$. Naturally, one may be interested in characterizing the center and dispersion of data on $\mathcal M$. For manifold data, the usual moments remain mathematically valid, but are difficult to interpret since they often fall outside $\mathcal M$. Therefore, it is useful to consider alternative notions of center and dispersion. We begin by defining a Bregman divergence.

Given a convex $\psi$ and a subgradient $z\in\partial\psi(y)$, the \emph{Bregman divergence} generated by $\psi$ is the gap at $u$ between $\psi$ and its linearization at $y$,
\(
    D^z_\psi(u,y) = \psi(u)-\psi(y)-\langle z,\,u-y\rangle,
\)
which is non-negative and vanishes when $u = y$ \citep{bregman1967relaxation}. When $\psi$ is differentiable at $y$, the subgradient is unique and the superscript $z$ may be dropped. When $\psi(u)=\|u\|^2/2$, the Bregman divergence is one half of the squared Euclidean distance, which one uses to define the center and dispersion for data in Euclidean space.

Generalizing this notion, let the center be the point of least average divergence from the data, and the dispersion be the average divergence that remains there. We have the Bregman center and dispersion:
    \(
        \mathcal C_\psi(z) = \arg\min_{w}\ \mathrm E\,D^{z}_\psi(w,\,y),\qquad
        \mathcal I_\psi(z) = \min_{w}\ \mathrm E\,D^{z}_\psi(w,\,y),
    \)
where the expectation is over the law of the latent $z: T(z) = y$, since $y$ is determined by $z$ but not conversely. It is not hard to see that $\psi(u) = \|u\|^2/2$ makes $T$ the identity, so that $z = y$, and recovers the Euclidean summaries $\mathcal C_\psi(z) = \mathrm E(y)$ and $\mathcal I_\psi(z) = \operatorname{tr}\operatorname{Cov}(y)/2$. Under the mirror likelihood both summaries have closed forms.

\begin{theorem}[Bregman center and dispersion under the mirror likelihood]
    \label{thm:bregman-moments}
    Let $z$ have mean $m = \mathrm E(z)$ and let $y = T(z)$, so that $z\in\mathcal F_{y} = \partial\psi(y)$. Then
    \begin{enumerate}
        \item[(i)] $T(m) \in \mathcal C_\psi(z)$;
        \item[(ii)] $\mathcal I_\psi(z) = \mathrm E\,\psi^*(z) - \psi^*(m) \ge 0$.
    \end{enumerate}
\end{theorem}

The above theorem is reminiscent of the common extrinsic approaches: the latent variables are already the mirror coordinates of the observations, so averaging them and mapping back gives the center.

\begin{remark}
    For higher-order moments, the mirror likelihood does not yield a simple closed form in general.
    Alternatively, for exponential-wrapped densities on a Riemannian manifold, \citet{li2020exponential} show that, writing $m^k_p = \mathrm E\{\log^{\mathcal M}_p(y)^{\otimes k}\}$ for the moments at a base point $p$, the change of variable via manifold-log map $z = \log^{\mathcal M}_p(y)$ identifies $m^k_p$ with the $k$th moment of the generating density on the tangent space $T_p\mathcal M$. Nevertheless, a restriction is that the change of variable requires $\log^{\mathcal M}_p$ to invert the exponential map, therefore the generating density for $z$ must be supported within the injectivity domain.
\end{remark}

\subsection{Models in correlation matrix space}

We now turn to the motivating problem of modeling correlation matrices. Compared to canonical settings of modeling covariance matrices, correlation matrices have the additional constraint of unit diagonal. Further, depending on the application, the correlation matrices may have further constraints. In this subsection, we discuss three common scenarios.

We first consider the case when the correlation matrix is strictly positive definite, which is common in applications when some preprocessing has been applied to the sample correlation, such as the \citet{ledoit2004well} smoothing popular in computational neuroscience. 
\begin{example}[Strictly positive definite correlation]
    \label{ex:correlation-matrices}
Let $\mathcal M = \{Y \in \mathbb S^d: Y \succ 0,\; \operatorname{diag}(Y) = \mathbf 1_d\}$ be the set of strictly positive definite (SPD) correlation matrices. Take the von Neumann entropy as the generator function,
\(
    \psi(W) = \operatorname{tr}(W\log W) + \iota_{\mathcal M}(W),
\)
where $\log$ is the matrix logarithm and the indicator enforces $Y_i\succ 0$ and unit diagonal. The gradient of $h(Y_i)=\operatorname{tr}(Y_i\log Y_i)$ on $\{Y_i\succ 0\}$ is
$
    \nabla h(Y_i) = \log Y_i + I.
$
The constraint $\operatorname{diag}(Y_i)=\mathbf 1_d$ consists of $d$ independent affine equalities. For $Y_i\in\mathcal M$, we have the normal cone $
    N_{\mathcal M}(Y_i) = \{\operatorname{diag}(v_i): v_i\in\mathbb R^d\}$,
corresponding to enforcing the $d$ affine equalities $\operatorname{diag}(Y_i)=\mathbf 1_d$.
Therefore, the fiber is
    \(\mathcal F_{Y_i} = \log Y_i + I + \{\operatorname{diag}(v_i): v_i \in\mathbb R^d\}.\)
Using independent Gaussian densities $Z_{i,jk}\sim\mathcal N(\mu_{jk},\sigma^2)$ for $k<j$ and $Z_{i,jj}\sim\mathcal N(\mu_{jj},\sigma^2/2)$ for $j=1,\dots,d$, and integrating out $v_i$, we obtain the marginal mirror likelihood:
\(
    \mathcal L \{ (Y_{i,jk})_{k<j}, (Z_{i,jk})_{k<j}; \theta\} \propto  \sigma^{-d(d-1)/2}\prod_{j=1}^d  
    \prod_{k<j} \exp\left [ -\frac{1}{2\sigma^2} \left\{ \left(\log Y_{i}\right)_{jk} - \mu_{jk} \right\}^2 \right].
\)
The likelihood involves only the strictly lower triangular part of $Z_i$, with $Y_i$ its deterministic transform; this is not completely surprising, since $(Y_{i})_{jj}=1$ is fixed.

Note that the mirror map from $Z_i$ to $Y_i$ would require solving
\(
   v_i: Y_i = \exp\!\bigl(Z_i - I - \operatorname{diag}(v_i)\bigr), \operatorname{diag}(Y_i)=\mathbf 1_d,
\)
which may not be straightforward. Nevertheless, since $Y_i$ is given, there is no need to solve for $v_i$; we need only reparameterize $Z_i$ in terms of $(Y_i,v_i)$. 
\exampleqed
\end{example}

Next, we consider the case when the correlation matrix is rank deficient and may have varying rank across the collection of correlation matrices. This is typical when the correlation matrix is estimated from a time series whose number of time points is smaller than the dimension of the matrix; further rank deficiency may occur when the time series data points are not linearly independent.

\begin{example}[Rank-deficient and varying correlation matrices]
    \label{ex:nuclear-unit-diag}
Let $\mathcal M = \{Y\in\mathbb S^d: Y\succeq 0,\ \operatorname{diag}(Y) = \mathbf 1_d\}$ be the elliptope of positive semidefinite matrices with unit diagonal, containing matrices of every rank from $1$ to $d$. We use $
    \psi(W) = \tfrac{1}{2}\|W\|_F^2 + \iota_{\mathcal M}(W)$, hence the mirror map is
\(
    T(Z) = \arg\max_{W\in\mathcal M}\bigl\{\langle Z,W\rangle - \tfrac{1}{2}\|W\|_F^2\bigr\}
    = \arg\min_{W\in\mathcal M}\tfrac{1}{2}\|W-Z\|_F^2,
\)
that is, the Frobenius projection of $Z$ onto $\mathcal M$, known as the nearest-correlation-matrix problem. The map is single-valued since the objective is strongly convex, and computable by alternating projections onto $\{W\succeq 0\}$ and $\{W_{ii}=1\}$ \citep{higham2002computing}, although again we do not need to compute the forward transform since $Y_i$ is given.

Since $h$ is smooth with $\nabla h(Y) = Y$, we just have to focus on the normal cone to the elliptope. Let $Y=U_+\Lambda_+ U_+^\top$ be the spectral decomposition with diagonal $\Lambda_+\succ 0$, and $U_0\in\mathbb{R}^{d\times d_0}:U_0^\top U_+=O$ be an orthonormal basis of the null space of $Y$. Using the fact that $\{W: \operatorname{diag}(W) = \mathbf 1_d\}$ and $\{W: W\succeq 0\}$ are both convex sets, and their relative interiors intersect, the normal cone of the intersection is the sum of the normal cones of the two sets,
\(
    N_{\mathcal M}(Y) = \bigl\{\operatorname{diag}(\nu):\ \nu\in\mathbb R^d\bigr\} + \bigl\{U_0 A U_0^\top:\ A\in\mathbb S^{d_0},\ A\preceq 0\bigr\}.
\)
Therefore, the fiber is
$
    \mathcal F_Y = Y + N_{\mathcal M}(Y).
$
With $Z_{i,jj}\sim\mathcal N(M_{jj},\sigma^2)$ and $Z_{i,jl}\sim\mathcal N(M_{jl},\sigma^2/2)$ independently for $j<l$, the diagonal entries can be integrated out as in the last example, leading to the marginal mirror likelihood:
\(
    \mathcal L \{ (Y_{i,jk})_{k<j}, (Z_{i,jk})_{k<j}; \theta\} \propto  \sigma^{-d(d-1)/2}\prod_{j=1}^d  
    \prod_{k<j} \exp\left [ -\frac{1}{2\sigma^2} \left\{ \left(Y_{i} + U_{0,i} A_i U_{0,i}^\top\right)_{jk} - M_{jk} \right\}^2 \right],
\)
where $U_{0,i}$ needs to be computed only once for each $i$, and $A_i$ is a free parameter to be updated in the posterior computation. Since we assign a Gaussian distribution to the elements of $Z_i$, $Z_i$ is full rank almost everywhere, hence $A_i\prec 0$ strictly and can be parameterized by  $A_i=-\exp(\tilde A_i)$ with an unconstrained symmetric $\tilde A_i\in \mathbb{S}^{d_0}$.
\exampleqed
\end{example}

The last scenario corresponds to the case when the sample correlation matrix is sparsified during pre-processing, corresponding to $Y_{i,jk}=0$ in some entries. This is typical in neuroscience applications when the correlation is expected to be zero between two regions of interest, hence the sparsity pattern is fixed and known a priori. This scenario also extends to precision matrix estimates from algorithms such as graphical lasso \citep{friedman2008sparse}, for which the sparsity pattern is not fixed and could vary individually.

\begin{example}[Sparse correlation matrices]
    \label{ex:sparse-correlation}
Let $\mathcal M = \{Y\in\mathbb S^d: Y\succeq 0,\ \operatorname{diag}(Y) = \mathbf 1_d\}$ be the elliptope and take the $\ell_1$ norm of the off-diagonal entries, with $\lambda_{jk}\ge 0$ for $j<k$,
\(
    \psi(W) = \tfrac{1}{2}\|W\|_F^2 + \sum_{j<k} \lambda_{jk}|W_{jk}| + \iota_{\mathcal M}(W),
\)
where $\lambda_{jk}$ is the sparsity penalty for the entry $W_{jk}$. 

For simplicity, if the sparsity pattern is known a priori, we set the sparsity penalty $\lambda_{jk}=\lambda_0>0$ for the known zero entries $(j,k)$, and $\lambda_{jk}=0$ otherwise. If the sparsity pattern is not known a priori, we set the sparsity penalty $\lambda_{jk}=\lambda_0$ for all $j<k$, since one could still obtain non-zero entries under the mirror map. In both cases, $\lambda_0$ is fixed.

The quadratic term keeps $\psi$ $1$-strongly convex, so the mirror map
\(
    T(Z) = \arg\min_{W\in\mathcal M}\Bigl\{\tfrac{1}{2}\|W-Z\|_F^2 + \sum_{j<k}\lambda_{jk}|W_{jk}|\Bigr\}
\)
is single-valued. The penalty is separable across entries, so its subdifferential is the entrywise set
\(
    \Gamma(Y) = \bigl\{S\in\mathbb S^d:\ S_{jj}=0;\ S_{jk} = \lambda_{jk}\operatorname{sign}(Y_{jk})\ \text{if}\ Y_{jk}\ne 0,\ S_{jk}\in[-\lambda_{jk},\lambda_{jk}]\ \text{if}\ Y_{jk}=0\bigr\},
\)
and with $U_0\in\mathbb R^{d\times d_0}$ and $A$ as in Example~\ref{ex:nuclear-unit-diag},
\(
    \mathcal F_Y = Y + \Gamma(Y) + \bigl\{\operatorname{diag}(\nu):\ \nu\in\mathbb R^d\bigr\} + \bigl\{U_0 A U_0^\top:\ A\in\mathbb S^{d_0},\ A\preceq 0\bigr\}.
\)
Using the same Gaussian distribution for $Z_{i,jk}$ as the last example, we have the marginal mirror likelihood:
\(
    &\mathcal L \{ (Y_{i,jk})_{k<j}, (Z_{i,jk})_{k<j}; \theta\} \propto \\
    & \sigma^{-d(d-1)/2}\prod_{j=1}^d  
    \prod_{k<j} \exp\left [ -\frac{1}{2\sigma^2} \left\{ \left(Y_{i}+ S_{i,jk} + U_{0,i} A_i U_{0,i}^\top\right)_{jk} - M_{jk} \right\}^2 \right],
\)
where, for an entry with $Y_{i,jk}=0$, the term $S_{i,jk}\in [-\lambda_{jk},\lambda_{jk}]$ can be viewed as a random shift that frees the entry from the zero constraint.
\exampleqed
\end{example}
In the last example, we consider the case when the matrix is sparse and potentially rank deficient. Another possibility is that the matrix is sparse but SPD. In that case, one can consider either a composite mirror map onto the SPD and the sparse matrix spaces, or a mixture of the von Neumann entropy and the $\ell_1$ norm as the generator.

\section{Consistency theory}

\subsection{Distributional posterior consistency}
\label{sec:distributional-posterior-consistency}

We first establish posterior consistency at the level of the induced distributions of the observations. Consider the latent regression model
$$
    z_i\mid x_i,\beta,\sigma^2
    \sim N_d(f(x_i,\beta),\sigma^2 I_d),
    \qquad
    y_i=T(z_i),
$$
where $x_i\in\mathbb R^p$ are fixed covariates, $\beta\in\mathbb R^{d\times p}$, $\sigma^2>0$, and $f:\mathbb{R}^p\times\mathbb{R}^{d\times p}\to\mathbb{R}^d$ is a known function, such as linear function. 

For $m\in\mathbb R^d$ and $s>0$, let $Q_{m,s}$ denote the distribution of $T(Z)$ when $Z\sim N_d(m,sI_d)$. Then $Y_i\mid x_i,\theta\sim Q_{f(x_i,\beta),\sigma^2}$ independently. Write $\theta=(\beta,\sigma^2)$ for the parameter, with parameter space $\Theta=\mathbb R^{d\times p}\times(0,\infty)$, and let $\theta_0=(\beta_0,\sigma_0^2)$ denote the true parameter. We assume independent priors $\beta_{ij}\stackrel{\mathrm{iid}}{\sim}N(0,\tau^2)$, and $\sigma^2\sim\operatorname{IGauss}(\mu_\sigma,\lambda_\sigma)$, where $\tau^2,\mu_\sigma,\lambda_\sigma>0$ are fixed. Let $\Pi$ denote the resulting
prior distribution. We write $P_{\theta}^{(n)}=\prod_{i=1}^n Q_{f(x_i,\beta),\sigma^2}$.

Define the average Hellinger distance
$$
    d_n^2(\theta,\theta')
    =
    \frac{1}{n}\sum_{i=1}^n
    h^2\!\left(
        Q_{f(x_i,\beta),\sigma^2},
        Q_{f(x_i,\beta'),{\sigma'}^2}
    \right),
$$
where $h^2(P,Q)=1-\int\sqrt{dP\,dQ}$. The quantity $d_n$ measures discrepancy between the induced conditional distributions and therefore does not require the parameterization $(\beta,\sigma^2)$ to be identifiable.

We establish distributional posterior consistency by applying Theorem~1 of \cite{choi2007posterior}, which gives sufficient conditions for posterior consistency under independent, non-identically distributed observations. We therefore verify the required prior mass and Kullback--Leibler conditions and construct exponentially consistent tests on suitable parameter sieves.

\begin{condition}[Uniform Lipschitz continuity]
\label{cond:uniform-lipschitz}
    There exists a constant $L_f<\infty$ such that
    for all $\beta,\beta'\in\mathbb R^{d\times p}$,
    $$
        \sup_{i\geq1}
        \|f(x_i,\beta)-f(x_i,\beta')\|_2
        \leq
        L_f\|\beta-\beta'\|_F.
    $$
\end{condition}
This condition ensures that changes in the regression parameter result in uniformly controlled changes in the latent means across all observations. 

For each observation, define
$$
    K_i(\theta_0,\theta)
    =
    K\!\left(
        Q_{f(x_i,\beta_0),\sigma_0^2},
        Q_{f(x_i,\beta),\sigma^2}
    \right),
$$
where $K(P,Q)$ denotes the Kullback--Leibler divergence from $P$ to $Q$, and
$$
    V_i(\theta_0,\theta)
    =
    \operatorname{Var}_{\theta_0}
    \left[
        \log
        \frac{dQ_{f(x_i,\beta_0),\sigma_0^2}}
             {dQ_{f(x_i,\beta),\sigma^2}}(Y_i)
    \right].
$$

The following lemma shows that the uniform Lipschitz continuity implies the Kullback--Leibler support requirement for the induced observation model.

\begin{lemma}[Kullback--Leibler support]
\label{lem:induced-kl-support}
    Suppose Condition~\ref{cond:uniform-lipschitz} holds. Then there exists a measurable set $B\subset\Theta$ with $\Pi(B)>0$ such that
    $$
        \sum_{i=1}^{\infty}
        \frac{V_i(\theta_0,\theta)}{i^2}<\infty
        \qquad
        \text{for every }\theta\in B,
    $$
    and, for every $\delta>0$,
    $$
        \Pi\!\left(
            B\cap
            \left\{
                \theta:
                \sup_{i\geq1}
                K_i(\theta_0,\theta)<\delta
            \right\}
        \right)>0.
    $$
\end{lemma}

We next construct an explicit sieve for the parameter. We equip the parameter space with the Euclidean norm $\|\theta-\theta'\|_2^2=\|\beta-\beta'\|_F^2+|\sigma^2-{\sigma'}^2|^2$. Furthermore, we denote by $B_\rho(\theta,\delta)$ the $\rho$-ball centered at $\theta$ with radius $\delta$, and by $N(\delta,\Theta,\rho)$ the corresponding covering number of $\Theta$.

We consider the parameter sieve $\Theta_n=\left\{(\beta,\sigma^2):\|\beta\|_F\le \sqrt n,\;\frac{1}{n}\le\sigma^2\le n\right\}$. For this sieve, define $L_n^2=\max\left\{\frac{L_f^2n}{8},\frac{c_dn^2}{8}\right\}$, where $c_d=\max\left\{1,\frac d2\right\}$.

\begin{lemma}[Sieve mass and entropy bounds]
\label{lem:sieve-bounds}
    There exist constants $C,c>0$ such that
    $$
        \Pi(\Theta_n^c)
        \leq
        Ce^{-cn}.
    $$
    Moreover, for every fixed $\epsilon>0$,
    $$
        \log
        N\left(
            \frac{\epsilon}{36L_n},
            \Theta_n,
            \|\cdot\|_2
        \right)
        \leq
        \frac{n\epsilon^2}{4}
    $$
    for all sufficiently large $n$.
\end{lemma}

The next lemma shows that this Euclidean covering property is sufficient for obtaining the Hellinger covers needed in the testing argument.

\begin{lemma}[Euclidean covers induce Hellinger covers]
\label{lem:euclidean-hellinger-cover}
    Suppose Condition~\ref{cond:uniform-lipschitz} holds. Then, for every
    $\theta,\theta'\in\Theta_n$,
    $$
        d_n(\theta,\theta')
        \leq
        L_n\|\theta-\theta'\|_2.
    $$
    Consequently, for every $r>0$,
    $$
        B_{\|\cdot\|_2}\left(
            \theta,\frac{r}{L_n}
        \right)\cap\Theta_n
        \subseteq
        B_{d_n}(\theta,r),
    $$
    and therefore
    $$
        N(r,\Theta_n,d_n)
        \leq
        N\left(
            \frac{r}{L_n},
            \Theta_n,
            \|\cdot\|_2
        \right).
    $$
\end{lemma}

The preceding lemma allows the Euclidean entropy bound in Lemma~\ref{lem:sieve-bounds} to be transferred directly to the metric $d_n$. We can therefore cover the Hellinger-separated part of the sieve by finitely many $d_n$-balls and combine pointwise Hellinger tests over this cover to obtain a test that is uniform over the entire alternative.

\begin{lemma}[Tests for Hellinger-separated induced laws]
\label{lem:distributional-hellinger-tests}
    Suppose Condition~\ref{cond:uniform-lipschitz} holds. For every $\epsilon>0$, there exist tests $\phi_n$ and constants $C_\epsilon,c_\epsilon>0$ such that
    $$
        \sum_{n=1}^{\infty}E_{\theta_0}[\phi_n]<\infty
    $$
    and
    $$
        \sup_{\substack{
            \theta\in\Theta_n\\
            d_n(\theta,\theta_0)\geq\epsilon
        }}
        E_\theta[1-\phi_n]
        \leq
        C_\epsilon e^{-c_\epsilon n}.
    $$
\end{lemma}

The preceding results verify the two main ingredients required by Theorem~1 of \cite{choi2007posterior}. We can therefore conclude posterior consistency in the metric $d_n$.

\begin{theorem}[Distributional posterior consistency]
\label{thm:distributional-posterior-consistency}
    Suppose Condition~\ref{cond:uniform-lipschitz} holds. Then, for every $\epsilon>0$,
    $$
        \Pi\!\left(
            d_n(\theta,\theta_0)\geq\epsilon
            \mid
            Y_1,\ldots,Y_n
        \right)
        \longrightarrow0
    $$
    $P_{\theta_0}^{(\infty)}$-almost surely.
\end{theorem}

Theorem~\ref{thm:distributional-posterior-consistency} concerns only recovery of the induced observation laws. Under additional conditions, this can be strengthened to posterior consistency for the model parameters themselves, which we consider next.

\subsection{Parameter recovery for the regression coefficient}
\label{sec:posterior-consistency}

We now strengthen distributional posterior consistency to consistency of the regression coefficient. For this purpose, we treat the latent variance as fixed and known, but retain the Gaussian prior on $\beta$ specified in the preceding subsection. Specifically, let $\sigma^2>0$ be fixed and consider
$$
    z_i\mid x_i,\beta
    \sim
    N_d(\beta x_i,\sigma^2 I_d),
    \qquad
    y_i=T(z_i).
$$

For $m\in\mathbb R^d$, let $Q_m$ denote the distribution of $T(Z)$ when $Z\sim N_d(m,\sigma^2I_d)$. Then $Y_i\mid x_i,\beta\sim Q_{\beta x_i}$ independently. We denote the true parameter by $\beta_0$ and write $P_{\beta}^{(n)}=\prod_{i=1}^n Q_{\beta x_i}$. Define $d_n^2(\beta,\beta')=\frac{1}{n}\sum_{i=1}^nh^2\!\left(Q_{\beta x_i},Q_{\beta' x_i}\right)$. 

\begin{condition}[Bounded covariates]
\label{cond:bounded-covariates}
    There exists $\ell<\infty$ such that
    $$
        \sup_{i\geq1}\|x_i\|_2\leq\ell.
    $$
\end{condition}

For the linear mean function $f(x_i,\beta)=\beta x_i$, Condition~\ref{cond:bounded-covariates} implies Condition~\ref{cond:uniform-lipschitz} with $L_f=\ell$. Consequently, the arguments of
Section~\ref{sec:distributional-posterior-consistency} also establish distributional posterior consistency for the present fixed-variance model. Although bounded covariates are not essential to the formulation of the model, such an assumption is commonly used as a technical regularity condition in the theoretical analysis of generalized linear models, particularly in the high-dimensional literature \citep{vandegeer2014, ningliu2017, xia2023}.

To strengthen this distributional consistency result to posterior consistency for $\beta$, it remains to show that parameters bounded away from $\beta_0$ are uniformly separated from the truth in the metric $d_n$. To establish this separation, we first derive two structural properties of the fixed-variance mirror location model.

\begin{lemma}[Monotonicity of the mirror map]
\label{lem:mirror-monotonicity}
    For any $z,z'\in\mathbb R^d$ at which $T$ is uniquely defined,
    \begin{equation}
    \label{eq:mirror-monotonicity}
        \left\langle
        z'-z,T(z')-T(z)
        \right\rangle
        \geq0.
    \end{equation}
    Moreover, equality holds only if $T(z)=T(z')$.
\end{lemma}

\begin{theorem}[Identifiability of the mirror location model]
\label{thm:mirror-identifiability}
    Define
    $$
        \mathcal V_{\mathcal M}
        =
        \operatorname{span}
        \left\{
        w-w':w,w'\in\mathcal M
        \right\}.
    $$
    For any $m,m'\in\mathbb R^d$,
    $$
        Q_m=Q_{m'}
        \quad\Longleftrightarrow\quad
        m-m'\in\mathcal V_{\mathcal M}^{\perp}.
    $$
    Consequently, $\{Q_m:m\in\mathbb R^d\}$ is identifiable if and only if $\operatorname{aff}(\mathcal M)=\mathbb R^d$.
\end{theorem}

We now impose the additional conditions required to translate distributional consistency into parameter consistency.

\begin{condition}[Mirror identifiability]
\label{cond:beta-mirror-identifiability}
    The feasible set satisfies $\operatorname{aff}(\mathcal M)=\mathbb R^d$.
\end{condition}

\begin{condition}[Minimum eigenvalue condition]
\label{cond:covariate-richness}
    There exists $\kappa>0$ such that, for all sufficiently large $n$,
    $$
        \lambda_{\min}\!\left(
            \frac{1}{n}
            \sum_{i=1}^n x_ix_i^\top
        \right)
        \geq\kappa.
    $$
\end{condition}

Condition~\ref{cond:beta-mirror-identifiability} ensures that distinct latent means induce distinct observation distributions when the latent variance is fixed. Together with bounded covariates, Condition~\ref{cond:covariate-richness} ensures that regression coefficients bounded away from the truth produce latent means bounded away from their true values for a nonvanishing fraction of the observations.

We first establish a directional separation argument for the fixed-variance Gaussian location family.

\begin{lemma}[Uniform directional separation]
\label{lem:directional-separation}
    Suppose Condition~\ref{cond:beta-mirror-identifiability} holds. Let $\mathcal K\subset\mathbb R^d$ be compact and let $a_0>0$. Then there exists $\Delta(\mathcal K,a_0)>0$ such that
    $$
        \inf_{\substack{
            m\in\mathcal K,\ \|v\|_2=1\\
            a\geq a_0
        }}
        \left\{
            E_{Q_{m+av}}
            \left[g(v^\top Y)\right]
            -
            E_{Q_m}
            \left[g(v^\top Y)\right]
        \right\}
        \geq
        \Delta(\mathcal K,a_0),
    $$
    where
    $$
        g(t)=\frac{1+\tanh(t)}{2}.
    $$
\end{lemma}

The next lemma converts separation of regression coefficients into separation of the induced observation distributions.

\begin{lemma}[Uniform regression separation]
\label{lem:beta-regression-separation}
    Suppose Conditions~\ref{cond:bounded-covariates}, \ref{cond:beta-mirror-identifiability}, and \ref{cond:covariate-richness} hold. Then, for every $\epsilon>0$, there exists $c_\epsilon>0$ such that
    $$
        \inf_{\|\beta-\beta_0\|_F\geq\epsilon}
        d_n(\beta,\beta_0)
        \geq
        c_\epsilon
    $$
    for all sufficiently large $n$.
\end{lemma}

We can now obtain posterior consistency of the regression coefficient directly from distributional posterior consistency.

\begin{theorem}[Posterior consistency of the regression coefficient]
\label{thm:mirror-regression-consistency}
    Suppose Conditions~\ref{cond:bounded-covariates}, \ref{cond:beta-mirror-identifiability}, and \ref{cond:covariate-richness} hold. Then, for every $\epsilon>0$,
    $$
        \Pi\!\left(
            \|\beta-\beta_0\|_F\geq\epsilon
            \mid
            Y_1,\ldots,Y_n
        \right)
        \longrightarrow0
    $$
    $P_{\beta_0}^{(\infty)}$-almost surely.
\end{theorem}

\section{Posterior computation}
\label{sec:computation}

Assigning a prior $\Pi_0(\theta)$ to the parameter of the latent density, the mirror likelihood gives the augmented posterior
\[\label{eq:augmented-posterior}
    \Pi(\theta, z_1,\dots,z_n\mid y_1,\dots,y_n) \propto \Pi_0(\theta)\prod_{i=1}^n \phi(z_i;\theta_i)\,\mathbf 1\bigl(z_i\in\mathcal F_{y_i}\bigr).
\]
Two features make \eqref{eq:augmented-posterior} convenient. First, the only $\theta$-dependent factor is the latent density $\phi$, which is a standard unconstrained model such as a Gaussian; there is no manifold normalizing constant depending on $\theta$. Second, the fiber constraint is separable across $i$ and free of $\theta$. Consequently, a data-augmentation Markov chain Monte Carlo scheme \citep{tanner1987calculation} alternates two blocks,
\(
    &\text{(i)}\quad z_i \sim \phi(z_i;\theta_i)\,\mathbf 1(z_i\in\mathcal F_{y_i}),\quad i=1,\dots,n,\\
    &\text{(ii)}\quad \theta \sim \Pi_0(\theta)\prod\nolimits_{i=1}^n \phi(z_i;\theta_i),
\)
and step (ii) is exactly the update one would use in an unconstrained model with $z_1,\dots,z_n$ observed. When $\phi$ is Gaussian and $\Pi_0$ is conjugate, one can update $\theta$ in closed form. We now focus on the update for step (i).

\subsection{Fiber with a closed-form reparameterization}
\label{sec:chart}

When the fiber admits a closed-form reparameterization, as in most examples in this article, we have $z_i = g(\eta_i)$ for an unconstrained fiber coordinate $\eta_i$ corresponding to $z_i\in\mathcal F_{y_i}$. We can sample in the space of $\eta_i$ using gradient-based algorithms, such as the Metropolis-adjusted Langevin algorithm (MALA) \citep{roberts1996exponential}. Since the sampler operates in the transformed space, the target must include the Jacobian factor, leading to 
\(
    \Pi(\eta_i\mid y_i,\theta_i) \propto  \phi(g(\eta_i);\theta_i)\,
    J_g(\eta_i).
\)
Here $J_g(\eta_i)=\sqrt{\det\!\left\{Dg(\eta_i)^\top Dg(\eta_i)\right\}}$ is the appropriate Jacobian when the transformation potentially changes dimension. Although the Jacobian factor is cumbersome to derive by hand, automatic differentiation obviates any manual calculation. Using MALA, the update for step (i) is then
\(
    \eta_i' = \eta_i + \tfrac{\varepsilon^2}{2}M\nabla_{\eta_i}\log\Pi(\eta_i\mid y_i,\theta_i) + \varepsilon\, e,
    \qquad e\sim \mathrm N(0, M),
\)
where $M$ is a positive definite matrix that can be chosen to precondition the gradient. Since the proposal is not symmetric, its density
$
    q(\eta_i'\mid \eta_i) = \mathrm N\Bigl(\eta_i';\ \eta_i + \tfrac{\varepsilon^2}{2}M\nabla_{\eta_i}\log\Pi(\eta_i\mid y_i,\theta_i),\ \varepsilon^2 M\Bigr)
$
enters the Metropolis--Hastings ratio, and $\eta_i'$ is accepted with probability
\(
    \min\left(1, \frac{\Pi(\eta_i'\mid y_i,\theta_i)\, q(\eta_i\mid \eta_i')}{\Pi(\eta_i\mid y_i,\theta_i)\, q(\eta_i'\mid \eta_i)}\right).
\)
Note that the proposal and acceptance are carried out individually for each $i$, so the acceptance probability is unaffected by the sample size $n$. Alternatively, under a conjugate prior, one could integrate out $\theta$ in step (ii) and update $\eta$ in the marginal posterior; however, this breaks the independence of the updates for $z_i$.

\subsection{Fiber with an intractable form}
\label{sec:ball-walk}

When no tractable reparameterization of $\mathcal F_{y_i}$ is available, we sample $z_i$ inside the fiber directly. By Theorem~\ref{thm:fiber-nonconvex} the fiber is a convex set, so a random walk on a convex body is applicable. We use the ball walk \citep{kannan2009random}: from the current $z_i$, propose a point $z_i'$ drawn uniformly from a ball of radius $\delta$ centered at $z_i$, discard the proposal if it falls outside $\mathcal F_{y_i}$, and otherwise apply a Metropolis--Hastings acceptance step for $z_i'$. However, the ball must be chosen with care. Although the fiber is convex, it is almost never full-dimensional in $\mathbb R^d$, so a ball taken in $\mathbb R^d$ would leave $\mathcal F_{y_i}$ with probability one, rejecting every proposal.

Therefore, the right ball should be taken inside the affine hull of the fiber. Since the subspace is spanned by differences of points of $\mathcal F_{y_i}$, as long as we can find enough points in $\mathcal F_{y_i}$, we can completely determine the subspace.

More specifically, by Theorem~\ref{thm:fiber-nonconvex}, $\mathcal F_{y_i}$ is the set of minimizers of
$
    F_{y_i}(z) = \psi^*(z) - \langle z, y_i\rangle.
$
One thing to note is that $\psi^*$ is always convex, regardless of the shape of $\psi$. We therefore obtain fiber points by minimizing $F_{y_i}$ from $K$ random starting values $r_1,\dots,r_K$, using the gradient-descent or subgradient method
$
    z^{t+1} = z^{t} - \alpha_t\bigl\{T(z^{t}) - y_i\bigr\},
$
which is stopped as soon as $T(z^{t}) = y_i$. Its optimal value $\min F_{y_i} = -\psi(y_i)$ is known in closed form, so the Polyak step size $\alpha_t = \{F_{y_i}(z^{t})+\psi(y_i)\}/\|T(z^{t})-y_i\|_2^2$ is available and no tuning is needed. Writing $z^{(k)}$ for the point returned from $r_k$, let $m_i$ denote the rank of
\(
    \bigl[z^{(2)}-z^{(1)},\ \dots,\ z^{(K)}-z^{(1)}\bigr],
\)
and let $V_i\in\mathbb R^{d\times m_i}$ collect its left singular vectors with non-zero singular values. Once $K$ is large enough, $m_i = \dim\operatorname{span}\bigl(\mathcal F_{y_i}-\mathcal F_{y_i}\bigr)$, the dimension of the affine hull of the fiber, and $V_i$ is an orthonormal basis of that span; in practice one raises $K$ until the rank stops growing, so that the span is not underestimated. Because $y_i$ is fixed, $V_i$ is computed once per observation before the chain starts, and it gives us a starting $z^{(1)}$ at the same time.

The ball-walk proposal is then
\(
    z_i' = z_i + \delta\, V_i u, \qquad u\sim \text{Uniform}\bigl(\mathbb B^{m_i}\bigr),
\)
with $\mathbb B^{m_i}$ the unit ball in $\mathbb R^{m_i}$. The walk moves $z_i$ inside that hull and rejects any proposal with $F_{y_i}(z_i')\ne -\psi(y_i)$.

Writing $z_i = z^{(1)}+V_ic_i$, the walk moves $c_i$ under the Lebesgue measure that the affine hull carries. One point needs care here. Since $y_i = T(z_i)$ is determined by $z_i$, the indicator in \eqref{eq:augmented-posterior} conditions on a set of measure zero, and a density on such a set is only defined once its dominating measure is named. Disintegrating the ambient Lebesgue measure along $z\mapsto T(z)$ gives $\mathrm dz = J(y,c)\,\mathrm dc\,\mathrm d\mu(y)$ for a reference measure $\mu$ on $\mathcal M$, so the target of the walk is
\(
    \Pi(c_i\mid y_i,\theta_i) \propto \phi\bigl(z^{(1)}+V_ic_i;\theta_i\bigr)\,J(y_i,c_i).
\)
The factor $J$ is the Jacobian of $(y_i,c_i)\mapsto z_i$. We adapt the step size so that the acceptance rate is close to $0.5$. As an alternative, hit-and-run \citep{smith1984efficient,belisle1993hit,lovasz1999hit} can be used in the same subspace; it avoids tuning $\delta$ at the cost of a line search for the boundary along each proposed direction.

\section{Simulation studies}

\subsection{Mirror regression on positive definite correlation matrices}
\label{sec:sim-consistency}

We take $\mathcal M$ and $\psi$ as in Example~\ref{ex:correlation-matrices} with $d=5$, so the ambient space is $\mathbb S^d$, and generate data from the mirror regression model of Section~\ref{sec:posterior-consistency}. The covariates are $x_i = (1,x_{i2},x_{i3})^\top\in\mathbb R^p$ with $p = 3$ and $x_{i2},x_{i3}\sim\mathrm N(0,1)$. Since the latent variable is a symmetric matrix, we attach one coefficient vector $\beta_{jk}\in\mathbb R^p$ to each entry $j\ge k$, so that
\(
    &Z_{i,jk}\mid x_i,\beta \sim \mathrm N\bigl(\beta_{jk}^\top x_i,\ \sigma^2\bigr)\ \ \text{for } k<j,
    \qquad
    Z_{i,jj}\mid x_i,\beta \sim \mathrm N\bigl(\beta_{jj}^\top x_i,\ \sigma^2/2\bigr),\\
    &Z_{i,kj} = Z_{i,jk},\qquad Y_i = T(Z_i),
\)
independently across $i$ and across entries, with $\sigma^2 = 0.09$; the two variances follow the convention of Example~\ref{ex:correlation-matrices}. Thus $\beta$ carries $d(d+1)/2$ coefficient vectors of length $p$, one per free entry of $Z_i$.

For the prior we assign $\beta_{jk,l}\mid\sigma^2\sim\mathrm N(0,\sigma^2\tau^2)$ independently for $k<j$ and $l = 1,\dots,p$, with $\tau = 10$, and $\sigma^2\sim\mathrm{IG}(10^{-3},10^{-3})$. The diagonal coefficients $\beta_{jj}$ are left out. Since every $Y_i$ has unit diagonal, the normal cone in Example~\ref{ex:correlation-matrices} leaves $\operatorname{diag}(Z_i)$ unconstrained along the fiber, so it integrates out of the marginal likelihood and the data carry no information about $\beta_{jj}$. This leaves the identified block $\{\beta_{jk}: k<j\}$, consisting of $\bar d = d(d-1)/2 = 10$ coefficient vectors and hence $\bar d p = 30$ scalar parameters, on which all errors below are computed.

\begin{figure}[H]
    \centering
    \includegraphics[width=\textwidth]{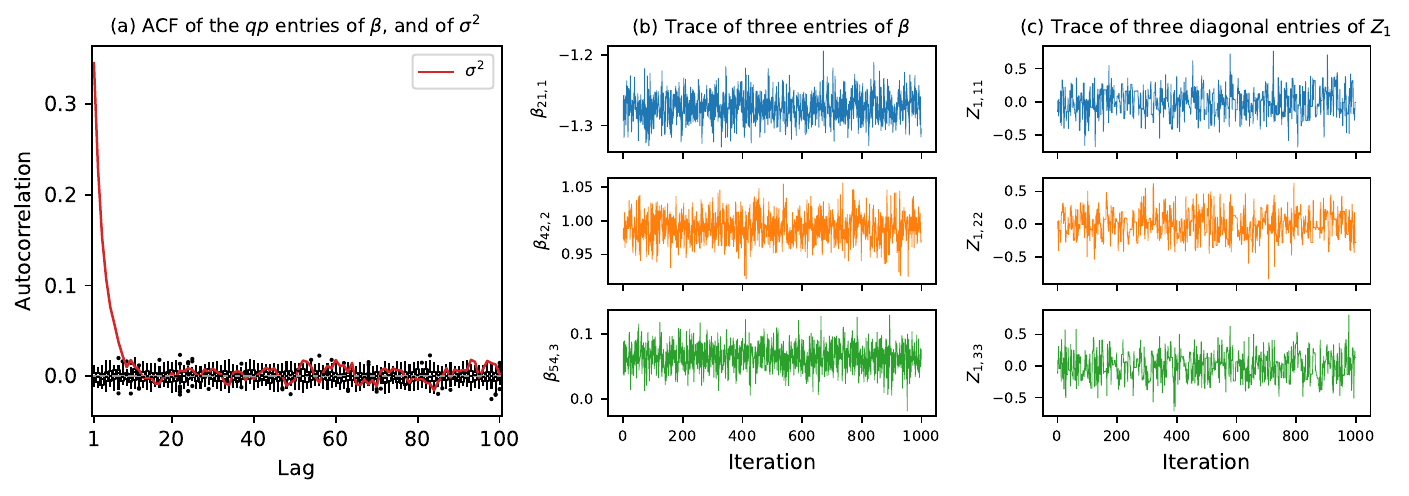}
    \caption{Sampler diagnostics for one chain of $20000$ iterations on a single data set with $n = 200$. (a) Boxplots, over the $\bar d p = 30$ entries of $\beta$, of the autocorrelation at each lag, with the autocorrelation of $\sigma^2$ overlaid. (b) Traces of three entries of $\beta$ over the first $1000$ iterations. (c) Traces of three diagonal entries of $Z_1$, the coordinates left free by the fiber.}
    \label{fig:mcmc-diagnostics}
\end{figure}

The sampler is the one of Section~\ref{sec:computation}: step (i) updates the free diagonal of each $Z_i$ by MALA, with proposals and acceptance carried out separately for each $i$, and step (ii) draws $\beta$ and then $\sigma^2$ from their conjugate full conditionals. To illustrate, we first take a single data set with $n = 200$ and run one chain for $20000$ iterations after $2000$ iterations of burn-in, with the MALA step size set by dual averaging \citep{hoffman2014no} to a target acceptance rate of $0.574$; the adapted step size is $1.29$ and the realized acceptance rate is $0.575$, and the run takes $2.3$ seconds. Figure~\ref{fig:mcmc-diagnostics} summarizes the output. The chain mixes rapidly: panel (a) shows that the autocorrelation of every entry of $\beta$ is indistinguishable from zero by lag one, while $\sigma^2$, which is shared across all $n$ observations, decays over roughly ten lags. Panels (b) and (c) show that the traces are stationary both for $\beta$ and for the diagonal entries of $Z_i$ that the fiber leaves free. The posterior mean of $\beta$ is within $0.02$ posterior standard deviations of its exact value at every one of the $30$ entries.

We next repeat the experiment over a range of sample sizes, with $200$ replicates on a nested design in which each replicate draws $n = 1600$ observations and the smaller sample sizes use the first $n$ of them. Table~\ref{tab:sim-consistency} reports the results. The posterior mean $\hat\beta$ satisfies $\sqrt n\,\|\hat\beta-\beta_0\|_F \approx 1.61$ across a $32$-fold range of $n$, indicating approximately $n^{-1/2}$ error decay; the coverage of marginal $95\%$ credible intervals, averaged over the $\bar d p = 30$ identified coordinates, stays within simulation error of the nominal level; and $\mathrm E(\sigma^2\mid Y_{1:n})$ approaches the true $0.09$.

\begin{table}[H]
    \centering
    \caption{Mirror regression on $5\times5$ correlation matrices, $200$ replicates. Errors are computed on the identified block of $\beta$, and coverage is averaged over its $\bar d p = 30$ entries.}
    \label{tab:sim-consistency}
    \begin{tabular}{rcccc}
        \toprule
        $n$ & $\|\hat\beta-\beta_0\|_F$ & $\sqrt n\,\|\hat\beta-\beta_0\|_F$ & Coverage & $\mathrm E(\sigma^2\mid Y_{1:n})$\\
        \midrule
        $50$   & $0.234$ & $1.653$ & $0.946$ & $0.0844$\\
        $100$  & $0.164$ & $1.640$ & $0.944$ & $0.0872$\\
        $200$  & $0.114$ & $1.614$ & $0.954$ & $0.0887$\\
        $400$  & $0.080$ & $1.605$ & $0.953$ & $0.0893$\\
        $800$  & $0.057$ & $1.611$ & $0.955$ & $0.0897$\\
        $1600$ & $0.040$ & $1.617$ & $0.951$ & $0.0899$\\
        \bottomrule
    \end{tabular}
\end{table}

\begin{figure}[H]
    \centering
    \includegraphics[width=\textwidth]{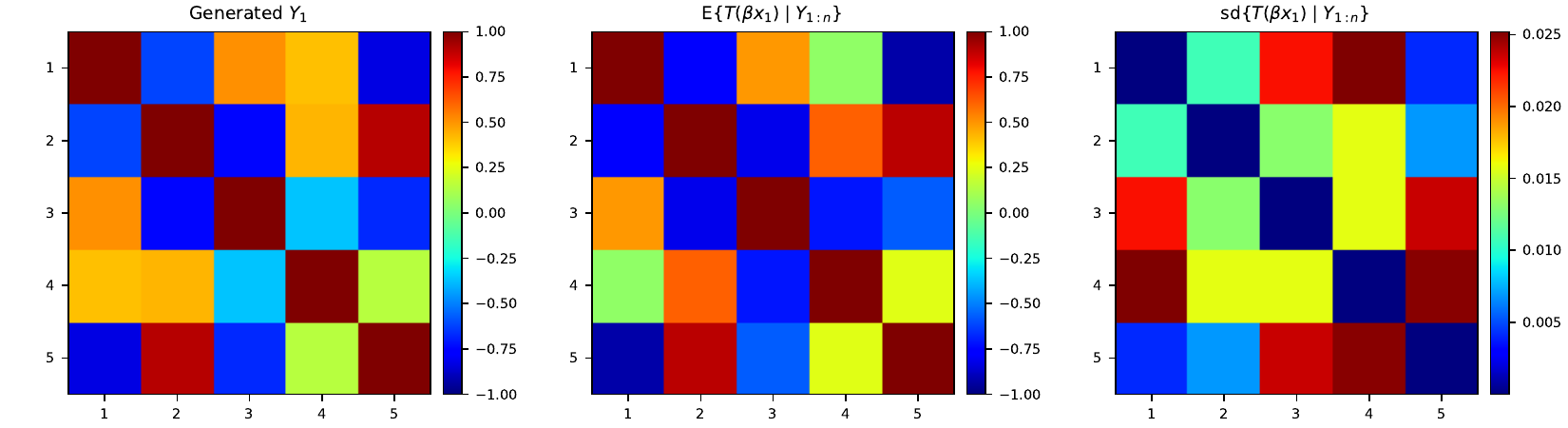}
    \caption{Left: one generated correlation matrix $Y_1$. Center and right: the posterior mean and posterior standard deviation of $T(\beta x_1)$. The left and center panels share a color scale, and entries that are exactly zero are drawn white.}
    \label{fig:mat-consistency}
\end{figure}

Figure~\ref{fig:mat-consistency} shows one generated correlation matrix alongside the posterior mean and standard deviation of $T(\beta x_1)$, the mirror map of the latent mean of $Z_1$. Applying $T$ puts the summary on $\mathcal M$, so it is directly comparable with the observation; note that $T$ is invariant to the diagonal of its argument, so the unidentified diagonal of $\beta$ does not obstruct this.

\subsection{Grassmann manifold}
\label{sec:sim-grassmann}
We now consider Example~\ref{ex:grassmann}, the Grassmann manifold of $k$-dimensional subspaces of $\mathbb R^d$, with $d = 5$ and $k = 2$. Its fiber is defined by the intractable inequality $\lambda_{\min}(A)\ge\lambda_{\max}(B)$, so we use the algorithm described in Section~\ref{sec:ball-walk}. The data are generated with a fixed mean $M\in\mathbb S^d$,
\(
    &Z_{i,jj}\mid M \sim \mathrm N\bigl(M_{jj},\ 1\bigr),
    \qquad
    Z_{i,jl}\mid M \sim \mathrm N\bigl(M_{jl},\ 1/2\bigr)\ \ \text{for } j<l,\\
    &Z_{i,lj} = Z_{i,jl},\qquad Y_i = T(Z_i),
\)
with $M$ carrying the eigenvalues $2$, $1.5$, $0$, $-1.5$ and $-2$ in a fixed randomly drawn basis, so that its leading two eigenvectors are separated from the rest. Here $T$ projects onto the top-$k$ eigenspace, so each $Y_i$ is a rank-two projection.

For the prior we take $M_{jl}\sim\mathrm N(0,\tau^2)$ with $\tau = 10$. To initialize the ball walk in the affine hull of the fiber, we minimize $F_{Y_i}(z) = \psi^*(z)-\langle z,Y_i\rangle$ from $16$ random starting values for each observation. The affine hull comes out at $m_i = 9$ for every observation, which is the intrinsic dimension of the fiber.

\begin{figure}[H]
    \centering
    \includegraphics[width=\textwidth]{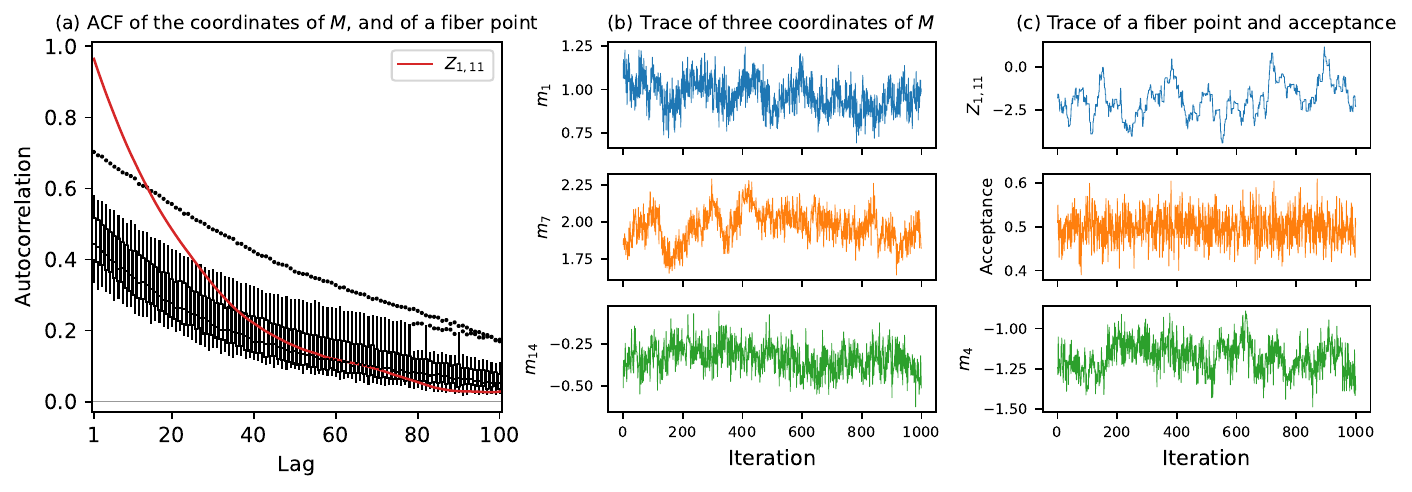}
    \caption{Sampler diagnostics for one chain of $20000$ iterations on a single data set with $n = 200$. (a) Boxplots, over the $\bar d = 14$ coordinates of $M$, of the autocorrelation at each lag, with the autocorrelation of one entry of a fiber point overlaid. (b) Traces of three coordinates of $M$ over the first $1000$ iterations. (c) Traces of that fiber entry, of the acceptance rate, and of a further coordinate of $M$.}
    \label{fig:grass-diagnostics}
\end{figure}

\begin{figure}[!htbp]
    \centering
    \includegraphics[width=\textwidth]{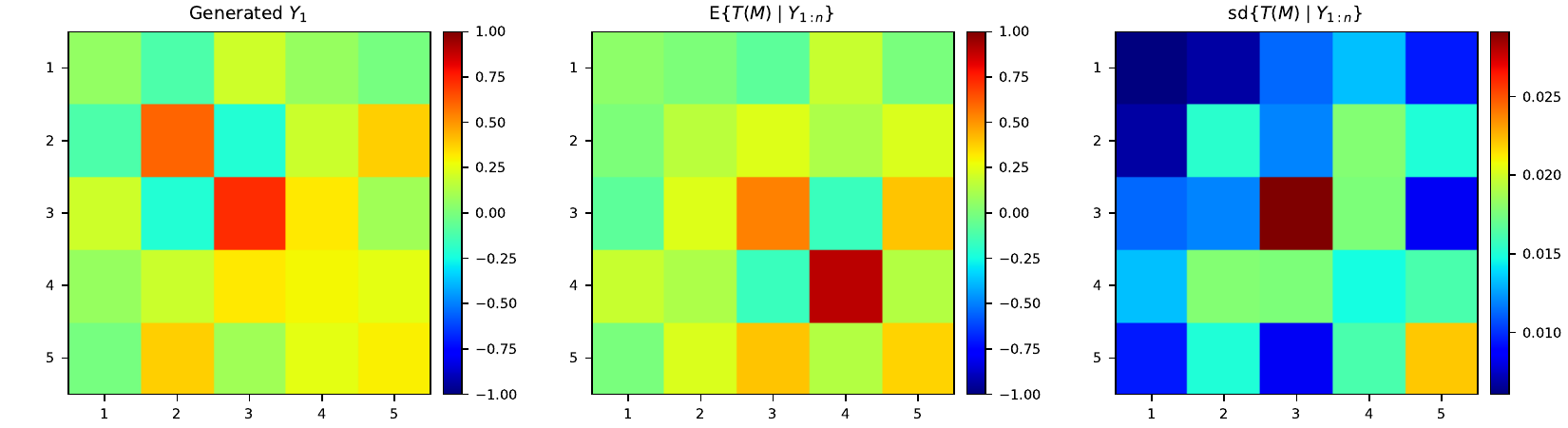}
    \caption{Left: one generated $Y_i$, a rank-two projection. Center and right: the posterior mean and posterior standard deviation of $T(M)$. The left and center panels share a color scale.}
    \label{fig:mat-grass}
\end{figure}

We run one chain of $20000$ iterations after $2000$ of burn-in, with the radius set by dual averaging to a target acceptance rate of $0.5$; the adapted radius is $\delta = 1.27$ and the realized acceptance rate is $0.499$. Figure~\ref{fig:grass-diagnostics} summarizes the output: the effective sample size of the $14$ coordinates of $M$ is at least $195$ with median $484$ out of $20000$ draws. Table~\ref{tab:sim-grass} reports the recovery of $M$, averaged over five replicates at each sample size. The scaled error $\sqrt n\,\|\hat M-M_0\|_F$ is flat between $4.75$ and $5.04$, so the error decays at the $n^{-1/2}$ rate, and the fitted subspace $T(\hat M)$ approaches $T(M_0)$. Figure~\ref{fig:mat-grass} shows one generated $Y_i$ alongside the posterior mean and standard deviation of $T(M)$. Notably, this posterior inference is carried out without explicitly calculating the intractable normalizing constant of the matrix Bingham distribution \citep{jupp1979maximum,hoff2009simulation}.

\begin{table}[H]
    \centering
    \caption{Grassmann manifold, one chain of $20000$ iterations per replicate, averaged over five replicates. Errors are computed on the $\bar d = 14$ identified coordinates of $M$.}
    \label{tab:sim-grass}
    \begin{tabular}{lccc}
        \toprule
        & $n = 200$ & $n = 800$ & $n = 3200$\\
        \midrule
        $\|\hat M-M_0\|_F$          & $0.356$ & $0.176$ & $0.084$\\
        $\sqrt n\,\|\hat M-M_0\|_F$ & $5.038$ & $4.978$ & $4.753$\\
        $\|T(\hat M)-T(M_0)\|_F$    & $0.088$ & $0.054$ & $0.012$\\
        \bottomrule
    \end{tabular}
\end{table}

\begin{figure}[H]
    \centering
    \includegraphics[width=\textwidth]{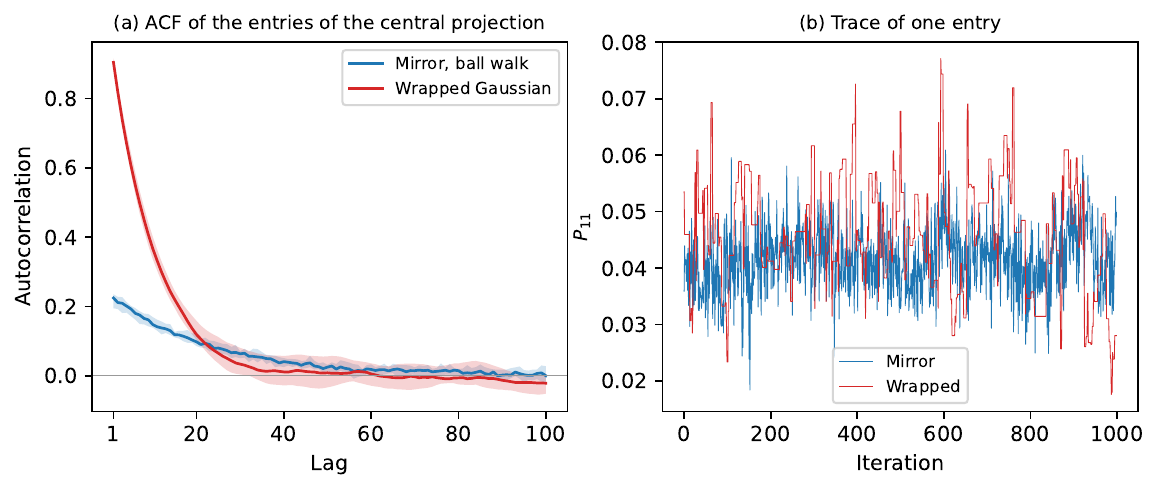}
    \caption{Mirror likelihood against a wrapped Gaussian on the same data set with $n = 200$. (a) Autocorrelation of the entries of the central projection, showing the median over entries with the range shaded. (b) Traces of one entry over the first $1000$ iterations.}
    \label{fig:wrapped-compare}
\end{figure}

For the Grassmann manifold, we can compare the performance with a tangent-space model available in closed form. In the wrapped Gaussian model, we start with a base point parameter $P\in\mathcal M$ and define $Y_i = \exp^{\mathcal M}_P(H_i)$, where $H_i\sim\mathrm N(0,\tau^2 I_q)$ and $q = k(d-k) = 6$. We place an inverse gamma prior on $\tau^2$ and integrate it out, and a uniform prior on $P$, so the chain only updates $P$ via a geodesic random walk, which we tune to achieve an acceptance rate of $0.234$. The two models have their center in different spaces, $M$ in the ambient space for the mirror likelihood and $P$ on $\mathcal M$ itself for the wrapped Gaussian, so we compare them on the central projection, $T(M)$ for the mirror likelihood and $P$ for the wrapped Gaussian. Figure~\ref{fig:wrapped-compare} shows that the mixing of the two models is comparable on the central projection. On the other hand, the mirror likelihood is more generally applicable, and can be used in settings where a tangent-space model is not available.

\section{Application to working-memory fMRI data}
\label{sec:fmri}

We apply the mirror likelihood to task functional magnetic resonance imaging (fMRI) data from the Human Connectome Project (HCP) \citep{vanessen2013wu,barch2013function}, with the goal of studying how the temporal evolution of functional connectivity differs with cognitive performance. We use the working-memory fMRI data and compare subjects in the lower and upper tails of the distribution of the Flanker score, a measure of inhibitory control and attention \citep{eriksen1974effects,weintraub2013cognition}.

The measurement obtained at each voxel is the blood-oxygen-level-dependent (BOLD) signal, the magnetic resonance contrast induced by the differing magnetic susceptibilities of oxygenated and deoxygenated hemoglobin \citep{ogawa1990brain}. Since neural activity is followed by a local increase in blood flow that outpaces oxygen consumption, the BOLD signal serves as an indirect, hemodynamically filtered proxy for neural activity, and is the standard readout in fMRI \citep{logothetis2008what}. For each subject, BOLD signals are averaged within $d=116$ anatomical regions of interest (ROIs) of the AAL parcellation \citep{tzourio2002automated}, giving 116 regional time series observed over 405 time points.

Individual BOLD time series carry substantial physiological and thermal noise at the single-time-point level, so the primary object of analysis in this literature is not the signal itself but the Pearson correlation matrix among regional series, treated as an estimate of functional connectivity. Computing these correlations within sliding windows yields dynamic functional connectivity, whose temporal variation is far better resolved than that of the raw signals because each window averages over many time points \citep{hutchison2013dynamic,allen2014tracking}. We construct dynamic functional connectivity using a sliding window of 110 observations, advanced by 10 observations at each step. The Pearson correlations among the 116 regional time series within each window give $Y_{it}\in\mathbb R^{116\times116}$, $t=1,\ldots,30$, where $Y_{it}\succeq0$ and $\operatorname{diag}(Y_{it})=\mathbf 1$.

Because each window contains fewer observations than there are ROIs, these matrices lie on the boundary of the elliptope rather than in its interior. In particular, a window of 110 time points caps the rank of $Y_{it}$ at 109, so the matrices in this analysis have nullity seven. The rank-deficient mirror likelihood of Example~\ref{ex:nuclear-unit-diag} therefore applies, with fiber dimension $d_{0,i}=7$ for every window.

We link the imaging data to the subjects' Flanker scores and retain subjects in the lower and upper 10\% of the score distribution. After matching the imaging and behavioral data, each group contains 104 subjects. We write $c=0$ and $c=1$ for the low- and high-Flanker groups, respectively.

Let $Y_{it}^{(c)}$ denote the correlation matrix for subject $i$, time $t$, and group $c$, and let $Z_{it}^{(c)}$ be its latent mirror coordinate. We model the temporal dependence through a bilinear matrix autoregressive model of order one \citep{chen2021autoregressive,basu2015regularized}:
$$
\begin{aligned}
    \widetilde Z_{it}^{(c)}
    &=
    A^{(c)}Z_{i,t-1}^{(c)}B^{(c)}
    +E_{it}^{(c)},\\
    Z_{it}^{(c)}
    &=
    S\!\left(\widetilde Z_{it}^{(c)}\right),
    && t=2,\ldots,30,\\
    \operatorname{vech}_L(E_{it}^{(c)})
    &\sim
    N_q(0,\sigma^2 I_q),
    && q=d(d+1)/2,
\end{aligned}
$$
where $S(\cdot)$ copies the lower-triangular entries, including the diagonal, to the upper triangle, and $\operatorname{vech}_L$ stacks the lower-triangular entries into a vector of length $q$. Conditional on the observed $Y_{it}^{(c)}$, the latent $Z_{it}^{(c)}$ is restricted to the fiber $\mathcal F_{Y_{it}^{(c)}}$. For the initial state, we use $\operatorname{vech}_L(Z_{i1}^{(c)}-Y_{i1}^{(c)})\sim N_q(0,\sigma^2 I_q)$, with $Z_{i1}^{(c)}\in\mathcal F_{Y_{i1}^{(c)}}$. The matrices $A^{(c)}$ and $B^{(c)}$ are shared across subjects within a group, while $\sigma^2$ is shared across the two groups.

We use Gaussian shrinkage priors centered at one for diagonal transition coefficients and zero for off-diagonal coefficients, with $A_{jj}^{(c)},B_{jj}^{(c)}\overset{\mathrm{ind}}{\sim}N(1,0.04)$ and $A_{jk}^{(c)},B_{jk}^{(c)}\overset{\mathrm{ind}}{\sim}N(0,0.01)$ for $j\neq k$, together with $\sigma^2\sim\operatorname{IG}(3,0.2)$. This is the Minnesota prior of the Bayesian vector autoregression literature \citep{doan1984forecasting,litterman1986forecasting}, which shrinks the transition towards a random walk and so guards against the overfitting that the $O(d^2)$ transition coefficients would otherwise invite.

The centering puts the prior mode at $A^{(c)}=B^{(c)}=I_d$, a random walk. That is the natural default here, as successive windows share 100 of their 110 time points. The diagonal scale $0.2$ then gives the self-persistence $A^{(c)}_{ii}B^{(c)}_{jj}$ of a connection a prior standard deviation of $0.29$, admitting both decay and mild amplification. The off-diagonal scale is half the diagonal one, the usual Minnesota downweighting of cross terms \citep{koop2010bayesian}. It still leaves room for cross-network propagation: a row of $A^{(c)}$ holds $115$ such coefficients, contributing $\sqrt{115}\times 0.1=1.07$, comparable to the own-lag term. Finally, the off-diagonal coordinates of $Z^{(c)}_{it}$ are of order one, so the prior mean $0.1$ for $\sigma^2$ sets the innovation scale an order of magnitude below the signal, and the shape $3$ is the smallest integer shape with finite variance. We experimented with higher-order autoregressions and found no practical difference, so we retain order one.

The bilinear mean admits the ambient vectorized representation $\operatorname{vec}(A^{(c)}ZB^{(c)})=(B^{(c)\top}\otimes A^{(c)})\operatorname{vec}(Z)$. In our model, however, only the lower-triangular entries are modeled directly, with the upper triangle determined by symmetry. Thus, writing $D_d$ and $L_d$ for the duplication and elimination matrices, respectively, the induced transition on the free symmetric coordinates is $K^{(c)}=L_d(B^{(c)\top}\otimes A^{(c)})D_d$. Both the ambient bilinear operator and $K^{(c)}$ are unchanged under the reciprocal rescaling $A^{(c)}\mapsto aA^{(c)}$ and $B^{(c)}\mapsto B^{(c)}/a$, whereas the individual factors $A^{(c)}$ and $B^{(c)}$ are not separately identified. To obtain a brain-level summary of this transition, consider an off-diagonal source connection $\{k,l\}$ with $k>l$. Since the corresponding free coordinate enters the symmetric latent matrix through both $Z_{kl}$ and $Z_{lk}$, its contribution to a future modeled connection $(i,j)$, $i>j$, is
$
    K_{(i,j),(k,l)}^{(c)}
    =
    A_{ik}^{(c)}B_{lj}^{(c)}
    +
    A_{il}^{(c)}B_{kj}^{(c)}.
$
We summarize the overall propagation of source connection $\{k,l\}$ by
$
    O_{kl}^{(c)}
    =
    \{
        \sum_{i>j}
        (
            K_{(i,j),(k,l)}^{(c)}
        )^2
    \}^{1/2},
    k>l.
$
Thus, $O_{kl}^{(c)}$ measures how strongly the previous connectivity between ROIs $k$ and $l$ propagates into the subsequent off-diagonal connectivity network. Because it depends only on products of entries of $A^{(c)}$ and $B^{(c)}$, it is invariant to their reciprocal scaling.

\begin{figure}[H]
    \centering
    \includegraphics[width=\textwidth]{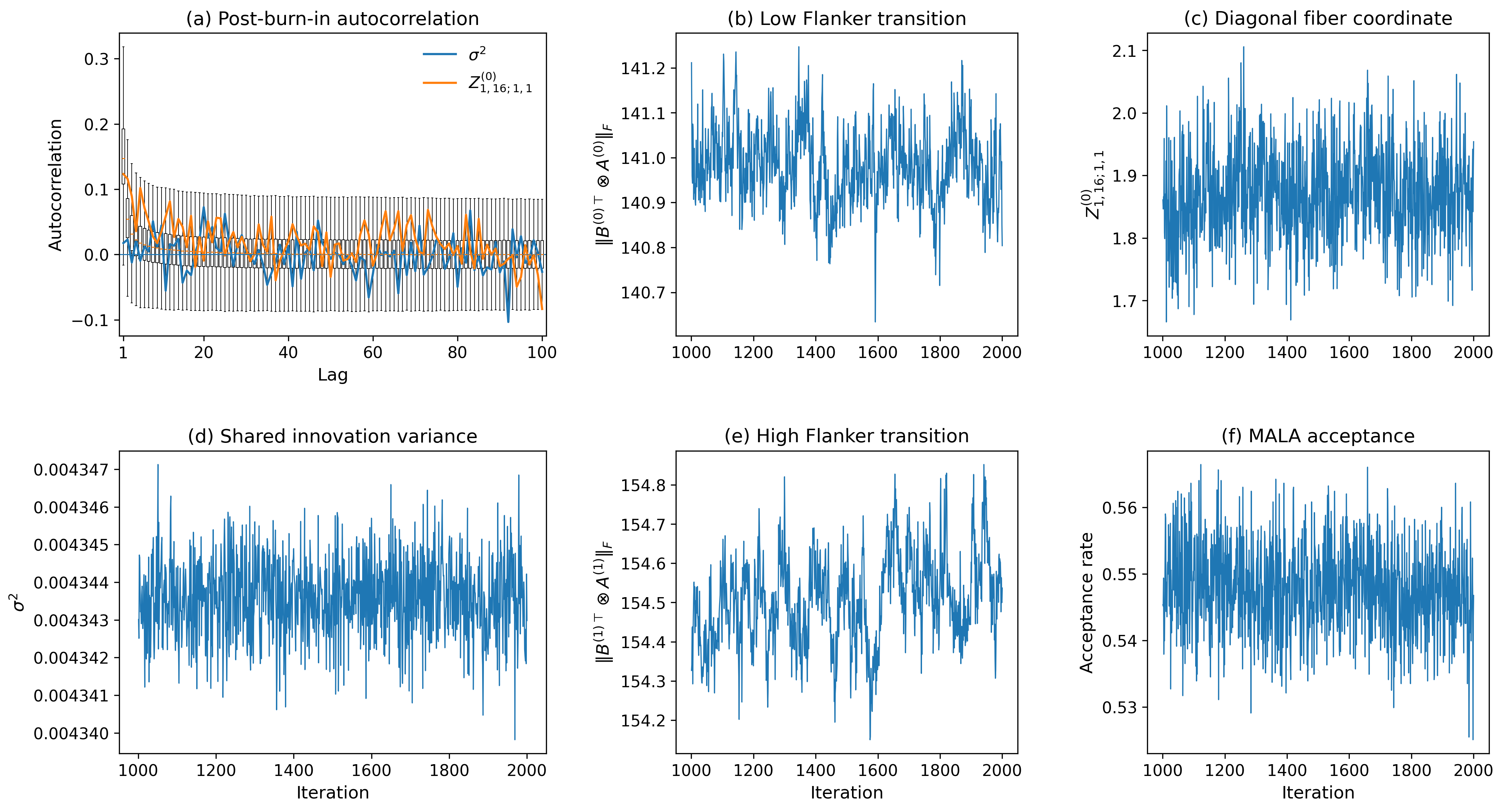}
    \caption{
    Post-burn-in MCMC diagnostics for the bilinear mirror autoregression. Panel (a) summarizes autocorrelations across entries of $A^{(c)}$ and $B^{(c)}$ together with $\sigma^2$ and a representative fiber coordinate. Panels (b) and (e) show traces of the ambient bilinear operator norm $\|B^{(c)\top}\otimes A^{(c)}\|_F$ for the low- and high-Flanker groups, respectively. Panel (c) shows a representative fiber coordinate, panel (d) shows $\sigma^2$, and panel (f) shows the MALA acceptance rate.
    }
    \label{fig:fmri-mcmc}
\end{figure}

We run the MCMC sampler for 2,000 iterations and discard the first 1,000 as burn-in. Figure~\ref{fig:fmri-mcmc} summarizes the post-burn-in mixing of the bilinear transition parameters, innovation variance, and representative fiber coordinates. The traces are stable and the autocorrelations decay rapidly, supporting use of the retained 1,000 draws for posterior comparison between groups.

The two groups differ substantially in their temporal propagation of functional connectivity. For posterior draw $r$, we summarize the overall off-diagonal propagation strength by $G_c^{(r)}=\{\sum_{k>l}(O_{kl}^{(c,r)})^2\}^{1/2}$. The high-Flanker group shows greater overall temporal propagation, with a posterior median difference of $11.87$ (Table~\ref{tab:fmri-source-influence}). The difference is also widespread across individual source connections. Among the 6670 possible ROI pairs, 97.1\% have a 95\% posterior credible interval for $\Delta O_{kl}=O_{kl}^{(1)}-O_{kl}^{(0)}$ that excludes zero.

To distinguish a change in network organization from a simple global increase in transition strength, we further define the relative source influence $R_{kl}^{(c,r)}=O_{kl}^{(c,r)}/G_c^{(r)}$. The between-group differences remain widespread after this normalization, with 97.6\% of source connections having 95\% credible intervals excluding zero. Thus, the high- and low-Flanker groups differ not only in the magnitude of temporal propagation, but also in how that influence is distributed across the network.

\begin{table}[H]
    \centering
    \caption{
    Posterior comparison of temporal source influence between the high- and low-Flanker groups.
    }
    \label{tab:fmri-source-influence}
    \begin{tabular}{lc}
        \hline
        Quantity & Posterior summary \\
        \hline
        Overall strength, Low Flanker
        & Median $=97.95$ \\

        Overall strength, High Flanker
        & Median $=109.82$ \\

        High $-$ Low overall strength
        & $11.87\;[11.68,\;12.04]$ \\

        Supported $\Delta O_{kl}$ connections
        & $6479/6670$ $(97.1\%)$ \\

        \quad High $>$ Low
        & $4369/6670$ $(65.5\%)$ \\

        \quad Low $>$ High
        & $2110/6670$ $(31.6\%)$ \\

        Supported normalized $\Delta R_{kl}$ connections
        & $6510/6670$ $(97.6\%)$ \\
        \hline
    \end{tabular}
\end{table}

\begin{figure}[H]
    \centering
    \includegraphics[width=\textwidth]{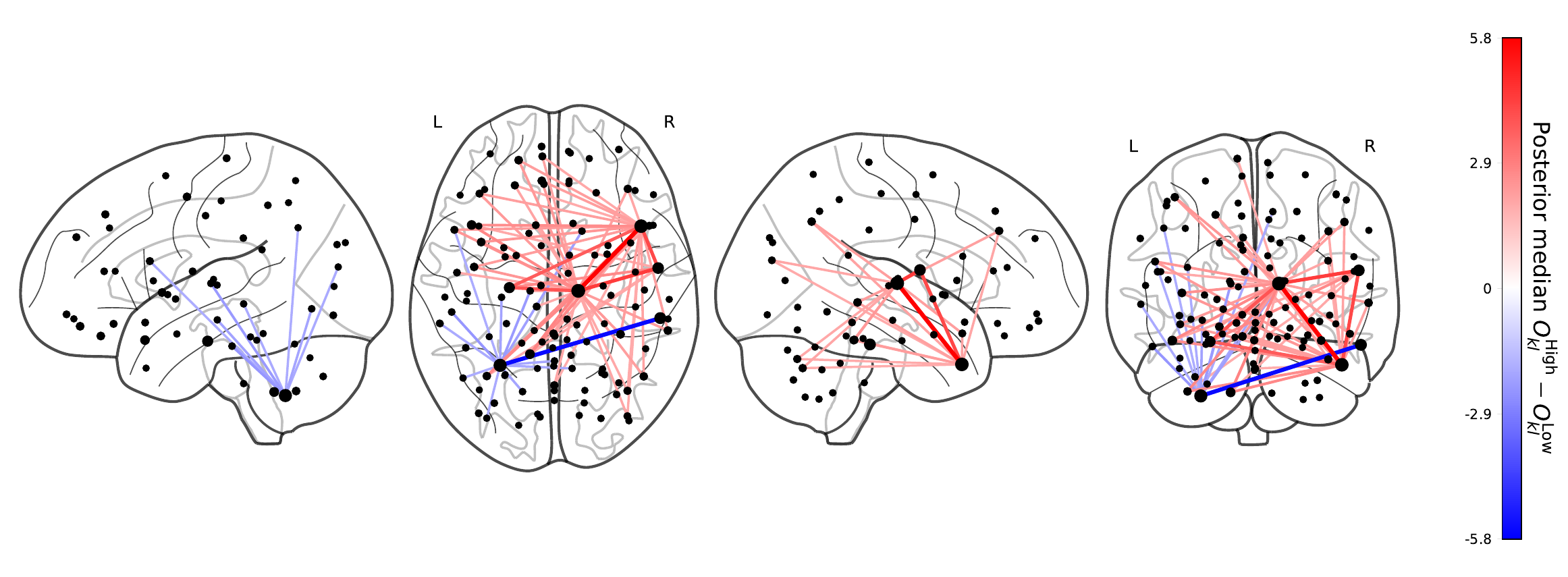}
    \caption{
    Differential temporal source influence between the high- and low-Flanker groups. Edges show the top 1\% of posterior-supported source connections ranked by the absolute posterior median of $\Delta O_{kl}=O_{kl}^{(1)}-O_{kl}^{(0)}$. Red indicates greater source influence in the high-Flanker group and blue indicates greater influence in the low-Flanker group. Node size is proportional to the sum of absolute High-minus-Low source-influence differences across all posterior-supported connections involving that ROI.
    }
    \label{fig:fmri-source-influence}
\end{figure}

Figure~\ref{fig:fmri-source-influence} shows where the strongest group differences occur. To avoid obscuring the network structure, we display only the top 1\% of posterior-supported source connections, ranked by $|\operatorname{median}(\Delta O_{kl})|$. Node size is proportional to the sum of absolute High-minus-Low source-influence differences across all posterior-supported connections involving that ROI.

The largest positive difference occurs for the connection between the right middle temporal pole and right thalamus, with posterior median $\Delta O=5.81$. Other prominent high-Flanker increases involve the right thalamus with the right rolandic operculum ($4.42$), and the right middle temporal pole with the right rolandic operculum ($4.18$). The strongest effect in the opposite direction is the connection between left cerebellar lobule VIII and the right inferior temporal gyrus, with posterior median $\Delta O=-5.59$. More broadly, the right thalamus and right middle temporal pole recur among the strongest high-Flanker effects, whereas left cerebellar lobule VIII features prominently among connections with stronger propagation in the low-Flanker group.

Together, these results suggest that Flanker performance is associated with both the magnitude and organization of the temporal propagation of working-memory functional connectivity. The high-Flanker group exhibits stronger propagation overall, while the persistence of widespread differences after normalization suggests that the groups differ not only in the overall magnitude of temporal influence, but also in how that influence is distributed across connections. The prominence of thalamic and cerebellar regions among the strongest differences is consistent with previous evidence implicating thalamocortical interactions and the posterior cerebellum in working-memory processing \citep{chen2023thalamocortical,salmi2010cognitive}.

\section{Discussion}
\label{sec:discussion}

We introduce the mirror map to enable likelihood-based inference for manifold-valued data. The framework offers three distinctive advantages. First, the augmented likelihood is tractable and carries no normalizing constant, and it can be read both as an extension of projected likelihood methods and as a sharper account of their statistical properties. Second, compared with tangent-space models, the construction imposes no stringent requirement on $\mathcal M$: under mild conditions on $h$, the space may be non-convex or non-smooth. The fiber is then always non-empty and convex, so every latent variable in Euclidean space can be read as the {\em mirror image} of a point on $\mathcal M$. Third, because the parameter enters only through the latent density, posterior computation reduces to standard unconstrained algorithms, as illustrated in the examples throughout the article.

Several directions are worth pursuing. First, for estimation and inference, we sample the posterior due to the close connection between augmented likelihood and the Bayesian framework; one could alternatively consider point estimation by the EM algorithm \citep{dempster1977maximum}. The complete-data likelihood is the latent density, so the E-step amounts to integrating the latent log-density over the fiber. Since the fiber is a convex body, this is a weighted-volume problem, and the randomized algorithms developed for approximating volumes of convex bodies \citep{dyer1991random,lovasz2006simulated,cousins2016practical} apply; when the fiber is polyhedral, the interior-point walks designed for polytopes \citep{chen2018fast} are a natural choice. A variational extension is also possible, replacing the exact E-step by a tractable surrogate that majorizes the objective, in the spirit of the majorization--minimization principle \citep{xu2017generalized,landeros2023mm}.

Second, our framework takes $\mathcal M$ as known and the observations as lying on it exactly, whereas a parallel literature treats the support itself as unknown or only partially known \citep{niu2019intrinsic,mukhopadhyay2020estimating,dunson2022graph,berenfeld2024estimating}. It would be interesting to explore the connection between those approaches and ours, and whether the two agree asymptotically as the data concentrate on $\mathcal M$. A natural step is to let the generator be learned rather than specified, which would extend the mirror likelihood to the case where the constraint is known only approximately.

\noindent\textbf{Disclosure of AI Assistance:}
We used Claude Code 5.5 for code writing and for polishing the manuscript. We take full responsibility for the content of the article.

\bibliographystyle{chicago}
\bibliography{references}

@article{roberts1996exponential,
author = {Gareth O. Roberts and Richard L. Tweedie},
title = {{Exponential convergence of Langevin distributions and their discrete approximations}},
volume = {2},
journal = {Bernoulli},
number = {4},
publisher = {Bernoulli Society for Mathematical Statistics and Probability},
pages = {341--363},
year = {1996},
}

@article{jauch2021monte,
  title={{M}onte {C}arlo simulation on the {S}tiefel manifold via polar expansion},
  author={Jauch, Michael and Hoff, Peter D and Dunson, David B},
  journal={Journal of Computational and Graphical Statistics},
  volume={30},
  number={3},
  pages={622--631},
  year={2021},
  publisher={Taylor \& Francis}
}

@article{khatri1977mises,
  title={The von {M}ises--{F}isher matrix distribution in orientation statistics},
  author={Khatri, CG and Mardia, Kanti V},
  journal={Journal of the Royal Statistical Society: Series B},
  volume={39},
  number={1},
  pages={95--106},
  year={1977},
  publisher={Oxford University Press}
}

@inproceedings{kannan2009random,
  title={Random walks on polytopes and an affine interior point method for linear programming},
  author={Kannan, Ravi and Narayanan, Hariharan},
  booktitle={Proceedings of the forty-first annual ACM symposium on Theory of computing},
  pages={561--570},
  year={2009}
}

@article{bhadra2024evidence,
  title={Evidence estimation in {G}aussian graphical models using a telescoping block decomposition of the precision matrix},
  author={Bhadra, Anindya and Sagar, Ksheera and Rowe, David and Banerjee, Sayantan and Datta, Jyotishka},
  journal={Journal of Machine Learning Research},
  volume={25},
  number={295},
  pages={1--43},
  year={2024}
}

@article{berger2020bayesian,
  title={Bayesian analysis of the covariance matrix of a multivariate normal distribution with a new class of priors},
  author={Berger, James O and Sun, Dongchu and Song, Chengyuan},
  journal={The Annals of Statistics},
  volume={48},
  number={4},
  pages={2381--2403},
  year={2020},
  publisher={JSTOR}
}

@article{lewandowski2009generating,
  title={Generating random correlation matrices based on vines and extended onion method},
  author={Lewandowski, Daniel and Kurowicka, Dorota and Joe, Harry},
  journal={Journal of Multivariate Analysis},
  volume={100},
  number={9},
  pages={1989--2001},
  year={2009},
  publisher={Elsevier}
}

@article{fisher1953dispersion,
  title={Dispersion on a sphere},
  author={Fisher, Ronald Aylmer},
  journal={Proceedings of the Royal Society of London. Series A},
  volume={217},
  number={1130},
  pages={295--305},
  year={1953}
}

@article{tanner1987calculation,
  title={The calculation of posterior distributions by data augmentation},
  author={Tanner, Martin A and Wong, Wing Hung},
  journal={Journal of the American Statistical Association},
  volume={82},
  number={398},
  pages={528--540},
  year={1987},
  publisher={Taylor \& Francis}
}

@book{mardia2000directional,
  title={Directional Statistics},
  author={Mardia, Kanti V and Jupp, Peter E},
  year={2000},
  publisher={Wiley}
}

@article{kendall1984shape,
  title={Shape manifolds, {P}rocrustean metrics, and complex projective spaces},
  author={Kendall, David G},
  journal={Bulletin of the London Mathematical Society},
  volume={16},
  number={2},
  pages={81--121},
  year={1984}
}

@book{dryden1998statistical,
  title={Statistical Shape Analysis},
  author={Dryden, Ian L and Mardia, Kanti V},
  year={1998},
  publisher={Wiley}
}

@article{alexander2007diffusion,
  title={Diffusion tensor imaging of the brain},
  author={Alexander, Andrew L and Lee, Jee Eun and Lazar, Mariana and Field, Aaron S},
  journal={Neurotherapeutics},
  volume={4},
  number={3},
  pages={316--329},
  year={2007}
}

@book{chikuse2003statistics,
  title={Statistics on Special Manifolds},
  author={Chikuse, Yasuko},
  year={2003},
  publisher={Springer}
}

@article{kolaczyk2017averages,
  title={Averages of unlabeled networks: Geometric characterization and asymptotic behavior},
  author={Kolaczyk, Eric D and Lin, Lizhen and Rosenberg, Steven and Walters, Jackson and Xu, Jie},
  journal={The Annals of Statistics},
  volume={48},
  number={1},
  pages={514--538},
  year={2020}
}

@article{ledoit2004well,
  title={A well-conditioned estimator for large-dimensional covariance matrices},
  author={Ledoit, Olivier and Wolf, Michael},
  journal={Journal of Multivariate Analysis},
  volume={88},
  number={2},
  pages={365--411},
  year={2004},
  publisher={Elsevier}
}

@article{hoff2009simulation,
  title={Simulation of the matrix {B}ingham--von {M}ises--{F}isher distribution, with applications to multivariate and relational data},
  author={Hoff, Peter D},
  journal={Journal of Computational and Graphical Statistics},
  volume={18},
  number={2},
  pages={438--456},
  year={2009},
  publisher={Taylor \& Francis}
}

@article{jupp1979maximum,
  title={Maximum likelihood estimators for the matrix von {M}ises--{F}isher and {B}ingham distributions},
  author={Jupp, Peter E and Mardia, Kanti V},
  journal={The Annals of Statistics},
  volume={7},
  number={3},
  pages={599--606},
  year={1979},
  publisher={Institute of Mathematical Statistics}
}

@book{bhattacharya2012nonparametric,
  title={Nonparametric Inference on Manifolds: With Applications to Shape Spaces},
  author={Bhattacharya, Abhishek and Bhattacharya, Rabindra N},
  year={2012},
  publisher={Cambridge University Press}
}

@article{bhattacharya2003large,
  title={Large sample theory of intrinsic and extrinsic sample means on manifolds},
  author={Bhattacharya, Rabindra N and Patrangenaru, Victor},
  journal={The Annals of Statistics},
  volume={31},
  number={1},
  pages={1--29},
  year={2003}
}

@article{bhattacharya2005large,
  title={Large sample theory of intrinsic and extrinsic sample means on manifolds--II},
  author={Bhattacharya, Rabindra N and Patrangenaru, Victor},
  journal={The Annals of Statistics},
  volume={33},
  number={3},
  pages={1225--1259},
  year={2005}
}

@article{pelletier2005kernel,
  title={Kernel density estimation on {R}iemannian manifolds},
  author={Pelletier, Bruno},
  journal={Statistics \& Probability Letters},
  volume={73},
  number={3},
  pages={297--304},
  year={2005}
}

@article{bhattacharya2010nonparametric,
  title={Nonparametric {B}ayesian density estimation on manifolds with applications to planar shapes},
  author={Bhattacharya, Abhishek and Dunson, David B},
  journal={Biometrika},
  volume={97},
  number={4},
  pages={851--865},
  year={2010}
}

@inproceedings{fletcher2011geodesic,
  title={Geodesic regression on {R}iemannian manifolds},
  author={Fletcher, P Thomas},
  booktitle={Proceedings of the Third International Workshop on Mathematical Foundations of Computational Anatomy},
  pages={75--86},
  year={2011}
}

@article{yuan2012local,
  title={Local polynomial regression for symmetric positive definite matrices},
  author={Yuan, Ying and Zhu, Hongtu and Lin, Weili and Marron, JS},
  journal={Journal of the Royal Statistical Society: Series B},
  volume={74},
  number={4},
  pages={697--719},
  year={2012}
}

@article{petersen2019frechet,
  title={Fr{\'e}chet regression for random objects with {E}uclidean predictors},
  author={Petersen, Alexander and M{\"u}ller, Hans-Georg},
  journal={The Annals of Statistics},
  volume={47},
  number={2},
  pages={691--719},
  year={2019}
}

@article{lin2017extrinsic,
  title={Extrinsic local regression on manifold-valued data},
  author={Lin, Lizhen and St Thomas, Brian and Zhu, Hongtu and Dunson, David B},
  journal={Journal of the American Statistical Association},
  volume={112},
  number={519},
  pages={1261--1273},
  year={2017}
}

@article{lin2019extrinsic,
  title={Extrinsic {G}aussian processes for regression and classification on manifolds},
  author={Lin, Lizhen and Mu, Niu and Cheung, Pokman and Dunson, David B},
  journal={Bayesian Analysis},
  volume={14},
  number={3},
  pages={887--906},
  year={2019}
}

@article{presnell1998projected,
  title={Projected multivariate linear models for directional data},
  author={Presnell, Brett and Morrison, Scott P and Littell, Ramon C},
  journal={Journal of the American Statistical Association},
  volume={93},
  number={443},
  pages={1068--1077},
  year={1998}
}

@article{wang2013directional,
  title={Directional data analysis under the general projected normal distribution},
  author={Wang, Fangpo and Gelfand, Alan E},
  journal={Statistical Methodology},
  volume={10},
  number={1},
  pages={113--127},
  year={2013}
}

@article{hernandez2017general,
  title={The general projected normal distribution of arbitrary dimension: Modeling and {B}ayesian inference},
  author={Hernandez-Stumpfhauser, Daniel and Breidt, F Jay and van der Woerd, Mark J},
  journal={Bayesian Analysis},
  volume={12},
  number={1},
  pages={113--133},
  year={2017}
}

@article{bregman1967relaxation,
  title={The relaxation method of finding the common point of convex sets and its application to the solution of problems in convex programming},
  author={Bregman, Lev M},
  journal={USSR Computational Mathematics and Mathematical Physics},
  volume={7},
  number={3},
  pages={200--217},
  year={1967}
}

@article{lin2017bayesian,
  title={{B}ayesian nonparametric inference on the {S}tiefel manifold},
  author={Lin, Lizhen and Rao, Vinayak and Dunson, David B},
  journal={Statistica Sinica},
  volume={27},
  number={2},
  pages={535--553},
  year={2017}
}

@article{stthomas2014learning,
  title={Learning subspaces of different dimension},
  author={St Thomas, Brian and Lin, Lizhen and Lim, Lek-Heng and Mukherjee, Sayan},
  journal={arXiv preprint arXiv:1404.6841},
  year={2014}
}

@article{wang2014modeling,
  title={Modeling space and space-time directional data using projected {G}aussian processes},
  author={Wang, Fangpo and Gelfand, Alan E},
  journal={Journal of the American Statistical Association},
  volume={109},
  number={508},
  pages={1565--1580},
  year={2014}
}

@article{nunez2015bayesian,
  title={{B}ayesian nonparametric models of circular variables based on {D}irichlet process mixtures of normal distributions},
  author={N{\'u}{\~n}ez-Antonio, Gabriel and Aus{\'\i}n, M Concepci{\'o}n and Wiper, Michael P},
  journal={Journal of Agricultural, Biological, and Environmental Statistics},
  volume={20},
  number={1},
  pages={47--64},
  year={2015}
}

@article{smith1984efficient,
  title={Efficient {M}onte {C}arlo procedures for generating points uniformly distributed over bounded regions},
  author={Smith, Robert L.},
  journal={Operations Research},
  volume={32},
  number={6},
  pages={1296--1308},
  year={1984}
}

@article{belisle1993hit,
  title={Hit-and-run algorithms for generating multivariate distributions},
  author={B{\'e}lisle, Claude J. P. and Romeijn, H. Edwin and Smith, Robert L.},
  journal={Mathematics of Operations Research},
  volume={18},
  number={2},
  pages={255--266},
  year={1993}
}

@article{friedman2008sparse,
  title={Sparse inverse covariance estimation with the graphical lasso},
  author={Friedman, Jerome H and Hastie, Trevor and Tibshirani, Robert},
  journal={Biostatistics},
  volume={9},
  number={3},
  pages={432--441},
  year={2008}
}

@article{lovasz1999hit,
  title={Hit-and-run mixes fast},
  author={Lov{\'a}sz, L{\'a}szl{\'o}},
  journal={Mathematical Programming},
  volume={86},
  number={3},
  pages={443--461},
  year={1999}
}

@article{higham2002computing,
  title={Computing the nearest correlation matrix---a problem from finance},
  author={Higham, Nicholas J.},
  journal={IMA Journal of Numerical Analysis},
  volume={22},
  number={3},
  pages={329--343},
  year={2002}
}

@article{choi2007posterior,
    title = {On posterior consistency in nonparametric regression problems},
    journal = {Journal of Multivariate Analysis},
    volume = {98},
    number = {10},
    pages = {1969--1987},
    year = {2007},
    author = {Taeryon Choi and Mark J. Schervish},
}

@article{ghosal2007convergence,
    title = {Convergence rates of posterior distributions for noniid observations},
    journal = {The Annals of Statistics},
    volume = {35},
    number = {1},
    pages = {192--223},
    year = {2007},
    author = {Ghosal, Subhashis and van der Vaart, Aad},
}

@article{hoffman2014no,
    title = {The {N}o-{U}-{T}urn sampler: Adaptively setting path lengths in {H}amiltonian {M}onte {C}arlo},
    journal = {Journal of Machine Learning Research},
    volume = {15},
    number = {1},
    pages = {1593--1623},
    year = {2014},
    author = {Hoffman, Matthew D. and Gelman, Andrew},
}

@article{li2020exponential,
    title = {Exponential-wrapped distributions on symmetric spaces},
    author = {Chevallier, Emmanuel and Li, Didong and Lu, Yulong and Dunson, David B.},
    journal = {SIAM Journal on Mathematics of Data Science},
    volume = {4},
    number = {4},
    pages = {1347--1368},
    year = {2022},
}

@article{vandegeer2014,
    title={On asymptotically optimal confidence regions and tests for high-dimensional models},
    volume={42},
    number={3},
    journal={The Annals of Statistics},
    pages={1166--1202},
    author={van de Geer, Sara and Bühlmann, Peter and Ritov, Ya’acov and Dezeure, Ruben},
    year={2014},
}

@article{ningliu2017,
  author  = {Ning, Yang and Liu, Han},
  title   = {A general theory of hypothesis tests and confidence regions for sparse high-dimensional models},
  journal = {The Annals of Statistics},
  year    = {2017},
  volume  = {45},
  number  = {1},
  pages   = {158--195},
}

@article{xia2023,
    author = {Xia, Lu and Nan, Bin and Li, Yi},
    title = {Debiased {L}asso for generalized linear models with a diverging number of covariates},
    journal = {Biometrics},
    volume = {79},
    number = {1},
    pages = {344--357},
    year = {2023},
}

@article{salmi2010cognitive,
    author = {Salmi, Juha and Pallesen, Karen Johanne and Neuvonen, Tuomas and Brattico, Elvira and Korvenoja, Antti and Salonen, Oili and Carlson, Synnöve},
    title = {Cognitive and motor loops of the human cerebro-cerebellar system},
    journal = {Journal of Cognitive Neuroscience},
    volume = {22},
    number = {11},
    pages = {2663--2676},
    year = {2010},
    month = {11},
}

@article{chen2023thalamocortical,
    title = {Thalamocortical contributions to working memory processes during the n-back task},
    journal = {Neurobiology of Learning and Memory},
    volume = {197},
    pages = {107701},
    year = {2023},
    author = {Xitong Chen and Evan Sorenson and Kai Hwang},
}

@article{vanessen2013wu,
    title = {The {WU-Minn} Human Connectome Project: An overview},
    author = {Van Essen, David C. and Smith, Stephen M. and Barch, Deanna M. and Behrens, Timothy E. J. and Yacoub, Essa and Ugurbil, Kamil},
    journal = {NeuroImage},
    volume = {80},
    pages = {62--79},
    year = {2013},
}

@article{barch2013function,
    title = {Function in the human connectome: Task-{fMRI} and individual differences in behavior},
    author = {Barch, Deanna M. and Burgess, Gregory C. and Harms, Michael P. and Petersen, Steven E. and Schlaggar, Bradley L. and Corbetta, Maurizio and Glasser, Matthew F. and Curtiss, Sandra and Dixit, Sachin and Feldt, Cindy and Nolan, Dan and Bryant, Edward and Hartley, Tucker and Footer, Owen and Bjork, James M. and Poldrack, Russell and Smith, Steve and Johansen-Berg, Heidi and Snyder, Abraham Z. and Van Essen, David C.},
    journal = {NeuroImage},
    volume = {80},
    pages = {169--189},
    year = {2013},
}

@article{ogawa1990brain,
    title = {Brain magnetic resonance imaging with contrast dependent on blood oxygenation},
    author = {Ogawa, Seiji and Lee, Tso-Ming and Kay, Alan R. and Tank, David W.},
    journal = {Proceedings of the National Academy of Sciences},
    volume = {87},
    number = {24},
    pages = {9868--9872},
    year = {1990},
}

@article{logothetis2008what,
    title = {What we can do and what we cannot do with {fMRI}},
    author = {Logothetis, Nikos K.},
    journal = {Nature},
    volume = {453},
    number = {7197},
    pages = {869--878},
    year = {2008},
}

@article{hutchison2013dynamic,
    title = {Dynamic functional connectivity: Promise, issues, and interpretations},
    author = {Hutchison, R. Matthew and Womelsdorf, Thilo and Allen, Elena A. and Bandettini, Peter A. and Calhoun, Vince D. and Corbetta, Maurizio and Della Penna, Stefania and Duyn, Jeff H. and Glover, Gary H. and Gonzalez-Castillo, Javier and Handwerker, Daniel A. and Keilholz, Shella and Kiviniemi, Vesa and Leopold, David A. and de Pasquale, Francesco and Sporns, Olaf and Walter, Martin and Chang, Catie},
    journal = {NeuroImage},
    volume = {80},
    pages = {360--378},
    year = {2013},
}

@article{allen2014tracking,
    title = {Tracking whole-brain connectivity dynamics in the resting state},
    author = {Allen, Elena A. and Damaraju, Eswar and Plis, Sergey M. and Erhardt, Erik B. and Eichele, Tom and Calhoun, Vince D.},
    journal = {Cerebral Cortex},
    volume = {24},
    number = {3},
    pages = {663--676},
    year = {2014},
}

@article{tzourio2002automated,
    title = {Automated anatomical labeling of activations in {SPM} using a macroscopic anatomical parcellation of the {MNI} {MRI} single-subject brain},
    author = {Tzourio-Mazoyer, N. and Landeau, B. and Papathanassiou, D. and Crivello, F. and Etard, O. and Delcroix, N. and Mazoyer, B. and Joliot, M.},
    journal = {NeuroImage},
    volume = {15},
    number = {1},
    pages = {273--289},
    year = {2002},
}

@article{eriksen1974effects,
    title = {Effects of noise letters upon the identification of a target letter in a nonsearch task},
    author = {Eriksen, Barbara A. and Eriksen, Charles W.},
    journal = {Perception \& Psychophysics},
    volume = {16},
    number = {1},
    pages = {143--149},
    year = {1974},
}

@article{weintraub2013cognition,
    title = {Cognition assessment using the {NIH} {Toolbox}},
    author = {Weintraub, Sandra and Dikmen, Sureyya S. and Heaton, Robert K. and Tulsky, David S. and Zelazo, Philip D. and Bauer, Patricia J. and Carlozzi, Noelle E. and Slotkin, Jerry and Blitz, David and Wallner-Allen, Kathleen and Fox, Nathan A. and Beaumont, Jennifer L. and Mungas, Dan and Nowinski, Cindy J. and Richler, Jennifer and Deocampo, Joanne A. and Anderson, Jacob E. and Manly, Jennifer J. and Borosh, Beth and Havlik, Richard and Conway, Kevin and Edwards, Emmeline and Freund, Lisa and King, Jonathan W. and Moy, Claudia and Witt, Ellen and Gershon, Richard C.},
    journal = {Neurology},
    volume = {80},
    number = {11, Supplement 3},
    pages = {S54--S64},
    year = {2013},
}

@article{chen2021autoregressive,
    title = {Autoregressive models for matrix-valued time series},
    author = {Chen, Rong and Xiao, Han and Yang, Dan},
    journal = {Journal of Econometrics},
    volume = {222},
    number = {1},
    pages = {539--560},
    year = {2021},
}

@article{basu2015regularized,
    title = {Regularized estimation in sparse high-dimensional time series models},
    author = {Basu, Sumanta and Michailidis, George},
    journal = {The Annals of Statistics},
    volume = {43},
    number = {4},
    pages = {1535--1567},
    year = {2015},
}

@article{litterman1986forecasting,
    title = {Forecasting with {B}ayesian vector autoregressions: Five years of experience},
    author = {Litterman, Robert B.},
    journal = {Journal of Business \& Economic Statistics},
    volume = {4},
    number = {1},
    pages = {25--38},
    year = {1986},
}

@article{doan1984forecasting,
    title = {Forecasting and conditional projection using realistic prior distributions},
    author = {Doan, Thomas and Litterman, Robert and Sims, Christopher},
    journal = {Econometric Reviews},
    volume = {3},
    number = {1},
    pages = {1--100},
    year = {1984},
}

@article{koop2010bayesian,
    title = {Bayesian multivariate time series methods for empirical macroeconomics},
    author = {Koop, Gary and Korobilis, Dimitris},
    journal = {Foundations and Trends in Econometrics},
    volume = {3},
    number = {4},
    pages = {267--358},
    year = {2010},
}

@article{dempster1977maximum,
    title = {Maximum likelihood from incomplete data via the {EM} algorithm},
    author = {Dempster, A. P. and Laird, N. M. and Rubin, D. B.},
    journal = {Journal of the Royal Statistical Society: Series B},
    volume = {39},
    number = {1},
    pages = {1--38},
    year = {1977},
}

@article{dyer1991random,
    title = {A random polynomial-time algorithm for approximating the volume of convex bodies},
    author = {Dyer, Martin and Frieze, Alan and Kannan, Ravi},
    journal = {Journal of the ACM},
    volume = {38},
    number = {1},
    pages = {1--17},
    year = {1991},
}

@article{lovasz2006simulated,
    title = {Simulated annealing in convex bodies and an {$O^*(n^4)$} volume algorithm},
    author = {Lov{\'a}sz, L{\'a}szl{\'o} and Vempala, Santosh},
    journal = {Journal of Computer and System Sciences},
    volume = {72},
    number = {2},
    pages = {392--417},
    year = {2006},
}

@article{cousins2016practical,
    title = {A practical volume algorithm},
    author = {Cousins, Ben and Vempala, Santosh},
    journal = {Mathematical Programming Computation},
    volume = {8},
    number = {2},
    pages = {133--160},
    year = {2016},
}

@article{chen2018fast,
    title = {Fast {MCMC} sampling algorithms on polytopes},
    author = {Chen, Yuansi and Dwivedi, Raaz and Wainwright, Martin J. and Yu, Bin},
    journal = {Journal of Machine Learning Research},
    volume = {19},
    number = {55},
    pages = {1--86},
    year = {2018},
}

@article{berenfeld2024estimating,
    title = {Estimating a density near an unknown manifold: A {Bayesian} nonparametric approach},
    author = {Berenfeld, Cl\'ement and Rosa, Paul and Rousseau, Judith},
    journal = {The Annals of Statistics},
    volume = {52},
    number = {5},
    pages = {2081--2111},
    year = {2024},
}

@article{dunson2022graph,
    title = {Graph based {Gaussian} processes on restricted domains},
    author = {Dunson, David B. and Wu, Hau-Tieng and Wu, Nan},
    journal = {Journal of the Royal Statistical Society: Series B},
    volume = {84},
    number = {2},
    pages = {414--439},
    year = {2022},
}

@article{niu2019intrinsic,
    title = {Intrinsic {Gaussian} processes on complex constrained domains},
    author = {Niu, Mu and Cheung, Pokman and Lin, Lizhen and Dai, Zhenwen and Lawrence, Neil and Dunson, David},
    journal = {Journal of the Royal Statistical Society: Series B},
    volume = {81},
    number = {3},
    pages = {603--627},
    year = {2019},
}

@article{mukhopadhyay2020estimating,
    title = {Estimating densities with non-linear support by using {Fisher--Gaussian} kernels},
    author = {Mukhopadhyay, Minerva and Li, Didong and Dunson, David B.},
    journal = {Journal of the Royal Statistical Society: Series B},
    volume = {82},
    number = {5},
    pages = {1249--1271},
    year = {2020},
}

@article{landeros2023mm,
    title = {{MM} optimization: Proximal distance algorithms, path following, and trust regions},
    author = {Landeros, Alfonso and Xu, Jason and Lange, Kenneth},
    journal = {Proceedings of the National Academy of Sciences},
    volume = {120},
    number = {27},
    pages = {e2303168120},
    year = {2023},
}

@inproceedings{xu2017generalized,
    title = {Generalized linear model regression under distance-to-set penalties},
    author = {Xu, Jason and Chi, Eric C. and Lange, Kenneth},
    booktitle = {Advances in Neural Information Processing Systems 30},
    pages = {1385--1395},
    year = {2017},
}

@article{sagar2024precision,
    title = {Precision matrix estimation under the horseshoe-like prior--penalty dual},
    author = {Sagar, Ksheera and Banerjee, Sayantan and Datta, Jyotishka and Bhadra, Anindya},
    journal = {Electronic Journal of Statistics},
    volume = {18},
    number = {1},
    pages = {1--46},
    year = {2024},
}

@article{lee2026scalable,
    title = {Scalable and optimal {B}ayesian inference for sparse covariance matrices via screened beta-mixture prior},
    author = {Lee, Kyoungjae and Jo, Seongil and Lee, Kyeongwon and Lee, Jaeyong},
    journal = {Bayesian Analysis},
    volume = {21},
    number = {2},
    pages = {541--568},
    year = {2026},
}

@article{bersson2025bayesian,
    title = {{B}ayesian covariance estimation for multi-group matrix-variate data},
    author = {Bersson, Elizabeth and Hoff, Peter D.},
    journal = {Bayesian Analysis},
    volume = {20},
    number = {4},
    pages = {1481--1507},
    year = {2025},
}

@article{yang2024empirical,
    title = {An empirical {B}ayes approach to shrinkage estimation on the manifold of symmetric positive-definite matrices},
    author = {Yang, Chun-Hao and Doss, Hani and Vemuri, Baba C.},
    journal = {Journal of the American Statistical Association},
    volume = {119},
    number = {545},
    pages = {259--272},
    year = {2024},
}

\clearpage
\section{Supplementary materials}

\subsection{Proofs}

\subsubsection*{Proofs for Section~\ref{subsec:fiber}}

\begin{proof}[Proof of Theorem~\ref{thm:subdifferential-guarantee}]
Since the result may be surprising to readers without an optimization background, we provide a short proof for the following equivalence:
\(
& y_i = \arg\max_w\{\langle z_i,w\rangle - \psi(w)\} \iff \psi(y_i)+\psi^*(z_i) = \langle z_i,y_i\rangle \\
& \iff \psi(w)\ge \psi(y_i)+\langle z_i, w-y_i\rangle \quad \forall w \in \mathbb R^d.
\)
The first equivalence is by definition of $\psi^*$. For the second equivalence, in the forward direction,  note that $\psi^*(z_i) = \sup_w\{\langle z_i,w\rangle-\psi(w)\} \ge \langle z_i,w\rangle - \psi(w) \forall w$, hence plugging it into the second equality gives 
\[\label{eq:non-convex-subdifferential}
\langle z_i,y_i\rangle - \psi(y_i) \;\ge\; \langle z_i,w\rangle - \psi(w)  \forall w,
\] which is equivalent to the last inequality. In the reverse direction, take supremum on the right-hand side of \eqref{eq:non-convex-subdifferential} to get $\psi^*(z_i) \le \langle z_i,y_i\rangle - \psi(y_i)$. But we know $\psi^*(z_i) \ge \langle z_i,w\rangle - \psi(w)$ for any $w$ including $w=y_i$, hence equality must hold.
\end{proof}

\begin{proof}[Proof of Theorem~\ref{thm:fiber-nonempty}]
    Let $\tilde h = h + \iota_C$, a proper convex function with $\operatorname{dom} \tilde h = C$. Since $y_i \in \operatorname{ri}(\operatorname{dom} \tilde h)$, we have $\partial \tilde h(y_i) \neq \varnothing$. Fix any $g \in \partial \tilde h(y_i)$. Because $\mathcal M \subseteq C$, we have $\psi \ge \tilde h$ everywhere, with equality at $y_i$, so
    $
    \psi(w)\;\ge\;\tilde h(w)\;\ge\;\tilde h(y_i)+\langle g,w-y_i\rangle\;=\;\psi(y_i)+\langle g,w-y_i\rangle\quad\forall w,
    $
    that is, $g\in \mathcal F_{y_i}$.
\end{proof}

\begin{proof}[Proof of Theorem~\ref{thm:fiber-nonconvex}]
    For each $w$, the linear inequality of $z_i$ corresponds to a half-space; the intersection of all such half-spaces is convex. For the second result, by Theorem~\ref{thm:subdifferential-guarantee}, $\psi^*(z)\ge \langle z,y_i\rangle - \psi(y_i)$ for every $z$,  with equality exactly on the fiber, hence given $y_i$, $\mathcal F_{y_i}$ is the set of minimizers.
\end{proof}

\subsubsection*{Proofs for Section~\ref{sec:center}}

\begin{proof}[Proof of Theorem~\ref{thm:bregman-moments}]
    Since $\psi^*(z) = \langle z,y\rangle - \psi(y)$, expanding the divergence and taking expectations, with $w$ non-random and $\mathrm E(z) = m$,
    \(
        \mathrm E\,D^{z}_\psi(w,y) = \psi(w) - \langle m,w\rangle + \mathrm E\,\psi^*(z).
    \)
    The last term is free of $w$, so minimizing is the same as maximizing $\langle m,w\rangle - \psi(w)$, which is \eqref{eq:fenchel_conjugate} at $m$; it is solved by $T(m)$ and attains $\psi^*(m)$. This gives (i), and substituting it gives the minimum value $\mathrm E\,\psi^*(z) - \psi^*(m)$, non-negative by Jensen's inequality since $\psi^*$ is convex.
\end{proof}

\subsubsection*{Proofs for Section~\ref{sec:distributional-posterior-consistency}}

\begin{proof}[Proof of Lemma~\ref{lem:induced-kl-support}]
    Set $m_{0i}=f(x_i,\beta_0)$ and $m_i=f(x_i,\beta)$. Since $Q_{m,s}$ is the pushforward of $N_d(m,sI_d)$ under the common measurable map $T$, the data-processing inequality gives
    $$
        K\!\left(
            Q_{m_{0i},\sigma_0^2},
            Q_{m_i,\sigma^2}
        \right)
        \leq
        K\!\left(
            N_d(m_{0i},\sigma_0^2I_d),
            N_d(m_i,\sigma^2I_d)
        \right)
        =
        \frac{d}{2}
        \left(
            \log\frac{\sigma^2}{\sigma_0^2}
            +\frac{\sigma_0^2}{\sigma^2}-1
        \right)
        +
        \frac{\|m_i-m_{0i}\|_2^2}{2\sigma^2}.
    $$
    By Condition~\ref{cond:uniform-lipschitz}, $\|m_i-m_{0i}\|_2 =\|f(x_i,\beta)-f(x_i,\beta_0)\|_2\leq L_f\|\beta-\beta_0\|_F$.
    Hence the Gaussian upper bound converges to zero uniformly in $i$ as $(\beta,\sigma^2)\to(\beta_0,\sigma_0^2)$. Therefore, for every $\delta>0$, there exists a neighborhood $\mathcal U_\delta$ of $\theta_0$ such that $\sup_{i\geq1}K_i(\theta_0,\theta)<\delta$ for every $\theta\in\mathcal U_\delta$.
    
    It remains to control $V_i(\theta_0,\theta)$. Let $L_i=\frac{dQ_{f(x_i,\beta_0),\sigma_0^2}}{dQ_{f(x_i,\beta),\sigma^2}}$. Using $(\log r)^2\leq r+r^{-1}$, for $r>0$, we obtain $V_i(\theta_0,\theta)\leq E_{\theta_0}\!\left[(\log L_i)^2\right]\leq E_{\theta_0}[L_i]+E_{\theta_0}[L_i^{-1}]$. Since $E_{\theta_0}[L_i^{-1}]=1$, while
    $$
        E_{\theta_0}[L_i]
        =
        \exp\!\left\{
            D_2\!\left(
                Q_{f(x_i,\beta_0),\sigma_0^2}
                \middle\|
                Q_{f(x_i,\beta),\sigma^2}
            \right)
        \right\},
    $$
    the data-processing inequality for the order-two R\'enyi divergence yields
    $$
        D_2\!\left(
            Q_{f(x_i,\beta_0),\sigma_0^2}
            \middle\|
            Q_{f(x_i,\beta),\sigma^2}
        \right)
        \leq
        D_2\!\left(
            N_d(f(x_i,\beta_0),\sigma_0^2I_d)
            \middle\|
            N_d(f(x_i,\beta),\sigma^2I_d)
        \right).
    $$
    
    For $P=N_d(m_0,\sigma_0^2I_d)$ and $Q=N_d(m,\sigma^2I_d)$, the order-two R\'enyi divergence is finite whenever $\sigma^2>\frac{\sigma_0^2}{2}$, and in that case
    $$
        D_2(P\|Q)
        =
        d\log\!\left(
            \frac{\sigma^2}
            {\sqrt{\sigma_0^2(2\sigma^2-\sigma_0^2)}}
        \right)
        +
        \frac{\|m-m_0\|_2^2}
             {2\sigma^2-\sigma_0^2}.
    $$
    
    Choose $\epsilon_0>0$ sufficiently small that $\epsilon_0<\frac{\sigma_0^2}{4}$, and define
    $$
        \mathcal U
        =
        \left\{
            (\beta,\sigma^2):
            \|\beta-\beta_0\|_F<\epsilon_0,
            \quad
            |\sigma^2-\sigma_0^2|<\epsilon_0
        \right\}.
    $$
    For every $\theta\in\mathcal U$, $\sigma^2>\sigma_0^2-\frac{\sigma_0^2}{4}=\frac{3\sigma_0^2}{4}$, and therefore $2\sigma^2-\sigma_0^2>\frac{\sigma_0^2}{2}>0$. Thus the Gaussian order-two R\'enyi divergence is finite throughout $\mathcal U$.
    
    Moreover, by Condition~\ref{cond:uniform-lipschitz}, $\|f(x_i,\beta)-f(x_i,\beta_0)\|_2\leq L_f\|\beta-\beta_0\|_F\leq L_f\epsilon_0$. Since $\sigma^2$ ranges over a compact interval contained in $(\sigma_0^2/2,\infty)$ on $\mathcal U$, it follows from the preceding formula that there exists $M<\infty$ such that
    $$\sup_{\theta\in\mathcal U}\sup_{i\geq1}D_2\!\left(N_d(f(x_i,\beta_0),\sigma_0^2I_d)\middle\|N_d(f(x_i,\beta),\sigma^2I_d)\right)\leq M.$$ Consequently, $V_i(\theta_0,\theta)\leq1+e^M$ uniformly over $\theta\in\mathcal U$ and $i\geq1$. Hence, writing $C=1+e^M$,
    $$
        \sum_{i=1}^{\infty}
        \frac{V_i(\theta_0,\theta)}{i^2}
        \leq
        C\sum_{i=1}^{\infty}\frac{1}{i^2}
        <\infty
    $$
    for every $\theta\in\mathcal U$.
    
    Choose $B=\mathcal U$. Since the Gaussian and inverse Gaussian prior densities are continuous and strictly positive at $\theta_0$, every neighborhood of $\theta_0$ has positive prior mass. In particular, $\Pi(B)>0$. Moreover, for every $\delta>0$, the set $B\cap\mathcal U_\delta$ is a neighborhood of $\theta_0$, and therefore $\Pi(B\cap\mathcal U_\delta)>0$. Since
    $$
        \mathcal U_\delta
        \subset
        \left\{
            \theta:
            \sup_{i\geq1}
            K_i(\theta_0,\theta)<\delta
        \right\},
    $$
    we obtain
    $$
        \Pi\!\left(
            B\cap
            \left\{
                \theta:
                \sup_{i\geq1}
                K_i(\theta_0,\theta)<\delta
            \right\}
        \right)>0.
    $$
    This proves the result.
\end{proof}

\begin{proof}[Proof of Lemma~\ref{lem:sieve-bounds}]
    The union bound gives
    $$
        \Pi(\Theta_n^c)
        \leq
        \Pi\left(\|\beta\|_F>\sqrt n\right)
        +
        \Pi\left(\sigma^2<\frac1n\right)
        +
        \Pi(\sigma^2>n).
    $$
    
    Since $\|\beta\|_F^2/\tau^2\sim\chi^2_{dp}$, we have $E[e^{t\|\beta\|_F^2}]=(1-2\tau^2t)^{-dp/2}$, for any $t<1/(2\tau^2)$. Thus, Markov's inequality gives $\Pi\left(\|\beta\|_F>\sqrt n\right)\leq(1-2\tau^2t)^{-dp/2}e^{-tn}$, for any fixed $t\in(0,1/(2\tau^2))$.

    Next, the inverse Gaussian density can be written as
    $$
        \pi(s)
        =
        C_0s^{-3/2}
        \exp\left\{
            -\frac{\lambda_\sigma}{2\mu_\sigma^2}s
            -\frac{\lambda_\sigma}{2s}
        \right\},
        \qquad s>0,
    $$
    for some constant $C_0>0$. Hence, for $n\geq1$,
    $$
    \begin{aligned}
        \Pi(\sigma^2>n)
        &\leq
        C_0\int_n^\infty
        \exp\left\{
            -\frac{\lambda_\sigma}{2\mu_\sigma^2}s
        \right\}\,ds \\
        &=
        C_1
        \exp\left\{
            -c_1n
        \right\},
    \end{aligned}
    $$
    for some $C_1,c_1>0$. Similarly,
    $$
    \begin{aligned}
        \Pi\left(\sigma^2<\frac1n\right)
        &\leq
        C_0\int_0^{1/n}
        s^{-3/2}
        \exp\left\{
            -\frac{\lambda_\sigma}{2s}
        \right\}\,ds \\
        &=
        C_0\int_n^\infty
        u^{-1/2}
        \exp\left\{
            -\frac{\lambda_\sigma}{2}u
        \right\}\,du \\
        &\leq
        C_2e^{-c_2n}
    \end{aligned}
    $$
    for some $C_2,c_2>0$, where we used the substitution $u=1/s$. Therefore, there exist constants $C,c>0$ such that $\Pi(\Theta_n^c)\leq Ce^{-cn}$ for all $n\geq1$.

    For the entropy bound, we have $L_n=O(n)$. Moreover, $\Theta_n$ is contained in a Euclidean
    ball in $\mathbb R^{dp+1}$ of radius $ R_n=\sqrt{n+n^2}=O(n)$. The standard covering bound for a Euclidean ball therefore gives
    $$
        N(\delta,\Theta_n,\|\cdot\|_2)
        \leq
        \left(
            1+\frac{2R_n}{\delta}
        \right)^{dp+1}.
    $$
    Taking $\delta=\epsilon/(36L_n)$,
    $$
    \begin{aligned}
        \log
        N\left(
            \frac{\epsilon}{36L_n},
            \Theta_n,
            \|\cdot\|_2
        \right)
        &\leq
        (dp+1)
        \log\left(
            1+\frac{72L_nR_n}{\epsilon}
        \right) \\
        &=O(\log n)
        =o(n).
    \end{aligned}
    $$
    Thus, for every fixed $\epsilon>0$,
    $$
        \log
        N\left(
            \frac{\epsilon}{36L_n},
            \Theta_n,
            \|\cdot\|_2
        \right)
        \leq
        \frac{n\epsilon^2}{4}
    $$
    for all sufficiently large $n$.
\end{proof}

\begin{proof}[Proof of Lemma~\ref{lem:euclidean-hellinger-cover}]
    Set $m_i=f(x_i,\beta)$ and $m_i'=f(x_i,\beta')$. Since $Q_{m,s}$ is the pushforward of $N_d(m,sI_d)$ under the common measurable map $T$, the data-processing inequality for Hellinger distance gives
    $$
        h\left(
            Q_{m_i,\sigma^2},
            Q_{m_i',{\sigma'}^2}
        \right)
        \leq
        h\left(
            N_d(m_i,\sigma^2I_d),
            N_d(m_i',{\sigma'}^2I_d)
        \right).
    $$

    The squared Hellinger distance between the two Gaussian laws is
    $$
        h^2\left(
            N_d(m_i,\sigma^2I_d),
            N_d(m_i',{\sigma'}^2I_d)
        \right)
        =
        1-
        \left(
            \frac{2\sqrt{\sigma^2{\sigma'}^2}}{\sigma^2+{\sigma'}^2}
        \right)^{d/2}
        \exp\left\{
            -\frac{\|m_i-m_i'\|_2^2}{4(\sigma^2+{\sigma'}^2)}
        \right\}.
    $$
    Using $1-ab\leq(1-a)+(1-b)$ for $a,b\in[0,1]$, we bound the mean and variance contributions separately.

    For the mean term, $1-e^{-u}\leq u$ and $\sigma^2+{\sigma'}^2\geq2/n$. Moreover, by Condition~\ref{cond:uniform-lipschitz}, $\|m_i-m_i'\|_2=\|f(x_i,\beta)-f(x_i,\beta')\|_2 \leq L_f\|\beta-\beta'\|_F$. Hence
    $$
        1-
        \exp\left\{
            -\frac{\|m_i-m_i'\|_2^2}{4(\sigma^2+{\sigma'}^2)}
        \right\}
        \leq
        \frac{L_f^2n}{8}
        \|\beta-\beta'\|_F^2.
    $$

    For the variance term, set $q=2\sqrt{\sigma^2{\sigma'}^2}/(\sigma^2+{\sigma'}^2)$. Since $0\leq q\leq1$, $1-q^{d/2}\leq c_d(1-q)$, where $c_d=\max\{1,d/2\}$. Using $\sigma^2,{\sigma'}^2\geq1/n$, it follows that
    $$
        1-q
        =
        \frac{(\sigma^2-{\sigma'}^2)^2}
        {(\sigma^2+{\sigma'}^2)(\sqrt{\sigma^2}+\sqrt{{\sigma'}^2})^2}
        \leq
        \frac{n^2(\sigma^2-{\sigma'}^2)^2}{8}.
    $$

    Combining these bounds with the data-processing inequality yields
    $$
        h^2\left(
            Q_{f(x_i,\beta),\sigma^2},
            Q_{f(x_i,\beta'),{\sigma'}^2}
        \right)
        \leq
        \frac{L_f^2n}{8}
        \|\beta-\beta'\|_F^2
        +
        \frac{c_dn^2}{8}
        |\sigma^2-{\sigma'}^2|^2.
    $$
    By the definition of $L_n$, the right-hand side is bounded by
    $L_n^2\|\theta-\theta'\|_2^2$. Averaging over $i=1,\ldots,n$
    therefore gives $d_n(\theta,\theta')\leq L_n\|\theta-\theta'\|_2$.

    Thus, if $\|\theta-\theta'\|_2<r/L_n$, then
    $d_n(\theta,\theta')<r$. Consequently,
    $$
        B_{\|\cdot\|_2}\left(
            \theta,\frac{r}{L_n}
        \right)\cap\Theta_n
        \subseteq
        B_{d_n}(\theta,r),
    $$
    and hence
    $$
        N(r,\Theta_n,d_n)
        \leq
        N\left(
            \frac{r}{L_n},
            \Theta_n,
            \|\cdot\|_2
        \right).
    $$
\end{proof}

\begin{proof}[Proof of Lemma~\ref{lem:distributional-hellinger-tests}]
    Fix $\epsilon>0$ and define $A_{\epsilon,n}=\{\theta\in\Theta_n:d_n(\theta,\theta_0)\geq\epsilon\}$. Let $r_\epsilon=\epsilon/18$ and $\eta_n=\epsilon/(36L_n)$. By Lemma~\ref{lem:sieve-bounds}, for all sufficiently large $n$, $\Theta_n$ can be covered by $N_n$ Euclidean balls of radius $\eta_n$, where $\log N_n\leq n\epsilon^2/4$. Retain only those balls that intersect $A_{\epsilon,n}$ and choose one point $\theta_{n,j}\in A_{\epsilon,n}$ from each such ball. If $\theta\in A_{\epsilon,n}$ belongs to the same Euclidean covering ball as $\theta_{n,j}$, then $\|\theta-\theta_{n,j}\|_2<2\eta_n$. By
    Lemma~\ref{lem:euclidean-hellinger-cover},
    $$
        d_n(\theta,\theta_{n,j})
        <
        2L_n\eta_n
        =
        \frac{\epsilon}{18}
        =:
        r_\epsilon.
    $$
    Hence $A_{\epsilon,n}$ can be covered by at most $N_n$ $d_n$-balls of radius $r_\epsilon$ whose centers belong to $A_{\epsilon,n}$. Since each center satisfies $d_n(\theta_{n,j},\theta_0)\geq\epsilon$, we have $r_\epsilon\leq d_n(\theta_{n,j},\theta_0)/18$.

    For each $j$, Lemma~2 of \cite{ghosal2007convergence} therefore
    provides a test $\phi_{n,j}$ such that
    $$
        E_{\theta_0}[\phi_{n,j}]
        \leq
        \exp\!\left\{
            -\frac{n}{2}
            d_n^2(\theta_{n,j},\theta_0)
        \right\}
        \leq
        e^{-n\epsilon^2/2},
    $$
    and
    $$
        \sup_{\theta\in
            B_{d_n}(\theta_{n,j},r_\epsilon)}
        E_\theta[1-\phi_{n,j}]
        \leq
        \exp\!\left\{
            -\frac{n}{2}
            d_n^2(\theta_{n,j},\theta_0)
        \right\}
        \leq
        e^{-n\epsilon^2/2}.
    $$
    By Lemma~\ref{lem:sieve-bounds}, there exists $N_\epsilon<\infty$ such that $\log N_n\leq n\epsilon^2/4$ for every $n\geq N_\epsilon$. For $n\geq N_\epsilon$, define the test for the full alternative by $\phi_n=\max_{1\leq j\leq N_n}\phi_{n,j}$, and for $n<N_\epsilon$, set $\phi_n\equiv0$.
    
    We first control the type-I error. For $n\geq N_\epsilon$, the union bound gives
    $$
        E_{\theta_0}[\phi_n]
        \leq
        \sum_{j=1}^{N_n}E_{\theta_0}[\phi_{n,j}]
        \leq
        N_n e^{-n\epsilon^2/2}
        \leq
        e^{-n\epsilon^2/4}.
    $$
    Since $\phi_n\equiv0$ for $n<N_\epsilon$, it follows that $\sum_{n=1}^{\infty}E_{\theta_0}[\phi_n]<\infty$.
    
    It remains to control the type-II error. For $n\geq N_\epsilon$ and $\theta\in A_{\epsilon,n}$, $\theta$ belongs to at least one covering ball, say $\theta\in B_{d_n}(\theta_{n,j},r_\epsilon)$. Since $\phi_n\geq\phi_{n,j}$,
    $$
        E_\theta[1-\phi_n]
        \leq
        E_\theta[1-\phi_{n,j}]
        \leq
        e^{-n\epsilon^2/2}.
    $$
    Therefore,
    $$
        \sup_{\theta\in A_{\epsilon,n}}
        E_\theta[1-\phi_n]
        \leq
        e^{-n\epsilon^2/2},
        \qquad n\geq N_\epsilon.
    $$
    
    For $n<N_\epsilon$, $\phi_n\equiv0$, so the type-II error is at most one. Thus, taking $c_\epsilon=\epsilon^2/2$ and $C_\epsilon=e^{c_\epsilon N_\epsilon}$,
    $$
        \sup_{\substack{
            \theta\in\Theta_n\\
            d_n(\theta,\theta_0)\geq\epsilon
        }}
        E_\theta[1-\phi_n]
        \leq
        C_\epsilon e^{-c_\epsilon n}
    $$
    for every $n\geq1$.
\end{proof}

\begin{proof}[Proof of Theorem~\ref{thm:distributional-posterior-consistency}]
    Fix $\epsilon>0$ and define $U_{n,\epsilon}=\left\{\theta\in\Theta: d_n(\theta,\theta_0)<\epsilon\right\}$. We apply Theorem~1 of \cite{choi2007posterior} with $U_n=U_{n,\epsilon}$.
    
    Lemma~\ref{lem:induced-kl-support} verifies condition (A1) of that
    theorem. Lemma~\ref{lem:distributional-hellinger-tests} provides tests
    satisfying conditions (A2)(i)--(ii) for $U_{n,\epsilon}^c\cap\Theta_n=\left\{\theta\in\Theta_n: d_n(\theta,\theta_0)\geq\epsilon\right\}$,
    and Lemma~\ref{lem:sieve-bounds} gives condition
    (A2)(iii).
\end{proof}

\subsubsection*{Proofs for Section~\ref{sec:posterior-consistency}}

\begin{proof}[Proof of Lemma~\ref{lem:mirror-monotonicity}]
    Let $y=T(z)$ and $y'=T(z')$. Optimality gives $\langle z,y\rangle-\psi(y)\geq\langle z,y'\rangle-\psi(y')$ and $\langle z',y'\rangle-\psi(y')\geq\langle z',y\rangle-\psi(y)$. Adding the inequalities yields \eqref{eq:mirror-monotonicity}.
    
    If equality holds, the two nonnegative optimality gaps must both vanish. Thus $y'$ is also optimal at $z$ and $y$ is also optimal at $z'$. Uniqueness at either point then gives $y=y'$.
\end{proof}

\begin{proof}[Proof of Theorem~\ref{thm:mirror-identifiability}]
    Let $v=m-m'$. If $v\in\mathcal V_{\mathcal M}^{\perp}$, then $\langle v,w\rangle$ is constant over $w\in\mathcal M$. Hence adding $v$ to the latent variable adds the same constant to every value of the optimization objective, so $T(z+v)=T(z)$ almost everywhere. Therefore $Q_{m'+v}=Q_{m'}$.
    
    Conversely, suppose $Q_m=Q_{m'}$ and let $Z\sim N_d(m', \sigma^2I_d)$, so that $Z+v\sim N_d(m,\sigma^2I_d)$. By Lemma~\ref{lem:mirror-monotonicity},
    $$
        \left\langle
        v,T(Z+v)-T(Z)
        \right\rangle
        \geq0
    $$
    almost surely. Since $Q_m=Q_{m'}$, $\langle v,T(Z+v)\rangle$ and $\langle v,T(Z)\rangle$ have the same distribution. Together with the almost-sure ordering, this implies equality almost surely. The equality case of Lemma~\ref{lem:mirror-monotonicity} and the almost-everywhere uniqueness of $T$ therefore give $T(Z+v)=T(Z)$ almost surely.
    
    Because the Gaussian density is strictly positive everywhere, $T(z+v)=T(z)$ for Lebesgue-almost every $z$. Taking countably many translations, there exists a full-measure set on which $T(z+kv)=T(z)$ for every $k\in\mathbb Z$.
    
    Fix such a $z$ and write $y=T(z)$. Since $y=T(z+kv)$ for every $k\in\mathbb Z$, optimality gives, for every $w\in\mathcal M$,
    $$
        k\langle v,y-w\rangle
        \geq
        \langle z,w-y\rangle+\psi(y)-\psi(w).
    $$
    Letting $k\to\infty$ and $k\to-\infty$ yields $\langle v,y-w\rangle=0$ for every $w\in\mathcal M$. Hence $v\in\mathcal V_{\mathcal M}^{\perp}$.
    
    Finally, $\mathcal V_{\mathcal M}=\mathbb R^d$ if and only if $\operatorname{aff}(\mathcal M)=\mathbb R^d$.
\end{proof}

\begin{proof}[Proof of Lemma~\ref{lem:directional-separation}]
    Define $\Delta(m,v,a)=E_{Q_{m+av}}[g(v^\top Y)]-E_{Q_m}[g(v^\top Y)]$. If $Z\sim N_d(m,\sigma^2I_d)$, Lemma~\ref{lem:mirror-monotonicity} gives
    $v^\top T(Z+av)\geq v^\top T(Z)$ almost surely. Since $g$ is increasing, $\Delta(m,v,a)\geq0$.
    
    Suppose $\Delta(m,v,a)=0$. The difference $g(v^\top T(Z+av))-g(v^\top T(Z))$ is nonnegative and has expectation zero, so it vanishes almost surely. Since $g$ is strictly increasing, $v^\top T(Z+av)=v^\top T(Z)$ almost surely. The equality case of Lemma~\ref{lem:mirror-monotonicity} therefore gives $T(Z+av)=T(Z)$ almost surely. Hence $Q_{m+av}=Q_m$. By Theorem~\ref{thm:mirror-identifiability} and Condition~\ref{cond:beta-mirror-identifiability}, this implies $av=0$. However, $a>0$ and $\|v\|_2=1$, so $av\neq0$, a contradiction. Therefore $\Delta(m,v,a)>0$.

    For fixed $a_0$, the map $(m,v)\mapsto\Delta(m,v,a_0)$ is continuous. To see this, let $(m_j,v_j)\to(m,v)$. Then $m_j+a_0v_j\to m+a_0v$, so the corresponding Gaussian measures converge in total variation. Since total variation contracts under measurable
    maps, $Q_{m_j+a_0v_j}\to Q_{m+a_0v}$ and $Q_{m_j}\to Q_m$ in total variation.
    
    Let $f_v(y)=g(v^\top y)$. For the first expectation, the triangle inequality gives
    $$
    \begin{aligned}
        \left|
        E_{Q_{m_j+a_0v_j}}[f_{v_j}(Y)]
        -
        E_{Q_{m+a_0v}}[f_v(Y)]
        \right|
        &\leq
        \left|
        E_{Q_{m_j+a_0v_j}}[f_{v_j}(Y)]
        -
        E_{Q_{m+a_0v}}[f_{v_j}(Y)]
        \right| \\
        &\quad+
        \left|
        E_{Q_{m+a_0v}}[f_{v_j}(Y)-f_v(Y)]
        \right|.
    \end{aligned}
    $$
    Since $0< f_{v_j}<1$, the variational characterization $d_{\mathrm{TV}}(P,Q)=\sup_{0\leq f\leq1}|E_Pf-E_Qf|$ implies that the first term is bounded by $d_{\mathrm{TV}}(Q_{m_j+a_0v_j},Q_{m+a_0v})$ and therefore converges to zero. For the second term, $v_j\to v$ implies $f_{v_j}(y)\to f_v(y)$ for every $y$, while $|f_{v_j}(y)-f_v(y)|\leq1$. Hence the dominated convergence theorem implies that the second term also converges to zero. The same argument gives $E_{Q_{m_j}}[f_{v_j}(Y)]\to E_{Q_m}[f_v(Y)]$. Therefore $\Delta(m_j,v_j,a_0)\to\Delta(m,v,a_0)$, proving continuity.
    
    Since $\mathcal K\times \{v\in\mathbb R^d:\|v\|_2=1\}$ is compact and $\Delta(m,v,a_0)>0$ everywhere on it,
    $$
        \Delta_0
        =
        \min_{\substack{m\in\mathcal K\\ \|v\|_2=1}}
        \Delta(m,v,a_0)
        >0.
    $$
    Finally, monotonicity along the ray implies $\Delta(m,v,a)\geq\Delta(m,v,a_0)$ whenever $a\geq a_0$. Taking $\Delta(\mathcal K,a_0)=\Delta_0$ proves the result.
\end{proof}

\begin{proof}[Proof of Lemma~\ref{lem:beta-regression-separation}]
    Fix $\beta$ with $\|\beta-\beta_0\|_F\geq\epsilon$ and write $\beta-\beta_0=tU$, where $t=\|\beta-\beta_0\|_F\geq\epsilon$ and $\|U\|_F=1$. Let $G_n=n^{-1}\sum_{i=1}^n x_ix_i^\top$. By Condition~\ref{cond:covariate-richness}, $G_n\succeq\kappa I_p$ for all sufficiently large $n$, and hence
    \begin{equation}
    \label{eq:design-separation}
        \frac1n\sum_{i=1}^n\|Ux_i\|_2^2
        =
        \operatorname{tr}(UG_nU^\top)
        \geq\kappa.
    \end{equation}
    
    Set $\eta=\sqrt{\kappa/2}$ and $I_n(U)=\{i:\|Ux_i\|_2\geq\eta\}$. Since $\|Ux_i\|_2\leq\ell$, while $\|Ux_i\|_2^2<\kappa/2$ for $i\notin I_n(U)$, \eqref{eq:design-separation} gives
    $$
        \kappa
        \leq
        \frac{|I_n(U)|}{n}\ell^2+\frac{\kappa}{2}.
    $$
    Therefore,
    $$
        \frac{|I_n(U)|}{n}
        \geq
        \frac{\kappa}{2\ell^2}
        =:\rho>0.
    $$
    
    For $i\in I_n(U)$, let $v_i=Ux_i/\|Ux_i\|_2$. Then $(\beta-\beta_0)x_i=t\|Ux_i\|_2v_i$ with $t\|Ux_i\|_2\geq\epsilon\eta=:a_0$.
    
    The true latent means $m_{0i}=\beta_0x_i$ belong to the compact set $\mathcal K_0=\{\beta_0x:\|x\|_2\leq\ell\}$. Hence Lemma~\ref{lem:directional-separation} yields $\Delta_\epsilon>0$ such that, for every $i\in I_n(U)$,
    $$
        E_{Q_{\beta x_i}}
        [g(v_i^\top Y_i)]
        -
        E_{Q_{\beta_0x_i}}
        [g(v_i^\top Y_i)]
        \geq
        \Delta_\epsilon.
    $$
    
    Because $0<g<1$, the variational characterization $d_{\mathrm{TV}}(P,Q)=\sup_{0\leq f\leq1}|E_Pf-E_Qf|$ implies $d_{\mathrm{TV}}(Q_{\beta x_i},Q_{\beta_0x_i})\geq\Delta_\epsilon$. Using $d_{\mathrm{TV}}(P,Q)\leq\sqrt2\,h(P,Q)$ gives
    $$
        h(Q_{\beta x_i},Q_{\beta_0x_i})
        \geq
        \frac{\Delta_\epsilon}{\sqrt2}
        =:b_\epsilon
    $$
    for every $i\in I_n(U)$. Therefore,
    $$
        d_n^2(\beta,\beta_0)
        \geq
        \frac{|I_n(U)|}{n}b_\epsilon^2
        \geq
        \rho b_\epsilon^2.
    $$
    Taking $c_\epsilon=\sqrt{\rho}\,b_\epsilon$ proves the result.
\end{proof}

\begin{proof}[Proof of Theorem~\ref{thm:mirror-regression-consistency}]
    Fix $\epsilon>0$. By Lemma~\ref{lem:beta-regression-separation}, there exists $c_\epsilon>0$ such that, for all sufficiently large $n$,
    $$
        \|\beta-\beta_0\|_F\geq\epsilon
        \quad\Longrightarrow\quad
        d_n(\beta,\beta_0)\geq c_\epsilon.
    $$
    Hence $\left\{\beta:\|\beta-\beta_0\|_F\geq\epsilon\right\}\subseteq\left\{\beta:d_n(\beta,\beta_0)\geq c_\epsilon\right\}$. Therefore,
    $$
        \Pi\!\left(
            \|\beta-\beta_0\|_F\geq\epsilon
            \mid
            Y_1,\ldots,Y_n
        \right) 
        \leq
        \Pi\!\left(
            d_n(\beta,\beta_0)\geq c_\epsilon
            \mid
            Y_1,\ldots,Y_n
        \right).
    $$
    The right-hand side converges to zero almost surely by Theorem~\ref{thm:distributional-posterior-consistency}, applied to the fixed-variance model. This proves the result.
\end{proof}

\subsection{Additional simulation studies}

\subsubsection*{Rank-deficient and rank-varying correlation matrices}

We take $\mathcal M$ and $\psi$ as in Example~\ref{ex:nuclear-unit-diag}, the elliptope with the squared Frobenius generator, again with $d = 5$, and a fixed mean $M\in\mathbb S^d$ in place of a regression:
\(
    &Z_{i,jk}\mid M \sim \mathrm N\bigl(M_{jk},\ \sigma^2\bigr)\ \ \text{for } k<j,
    \qquad
    Z_{i,jj}\mid M \sim \mathrm N\bigl(M_{jj},\ \sigma^2/2\bigr),\\
    &Z_{i,kj} = Z_{i,jk},\qquad Y_i = T(Z_i),
\)
independently across $i$ and across entries, with $\sigma^2 = 0.09$ and $T$ the nearest-correlation-matrix map, evaluated by Dykstra's alternating projections \citep{higham2002computing}. Taking $M$ to be an equicorrelation matrix with correlation $0.6$ modified in three entries, the $n = 200$ observations come out with ranks $5$, $4$, $3$ and $2$ in counts $14$, $126$, $58$ and $2$, so the fiber dimension $d_{0,i} = d - \operatorname{rank}(Y_i)$ varies across observations. The prior is as in Section~\ref{sec:sim-consistency}, with $\tau = 10$; the diagonal of $Z_i$ is free through $\nu_i$, so $M_{jj}$ is again unidentified and is left out, leaving the $\bar d = 10$ entries $M_{jk}$ with $k<j$.

\begin{figure}[!htbp]
    \centering
    \includegraphics[width=\textwidth]{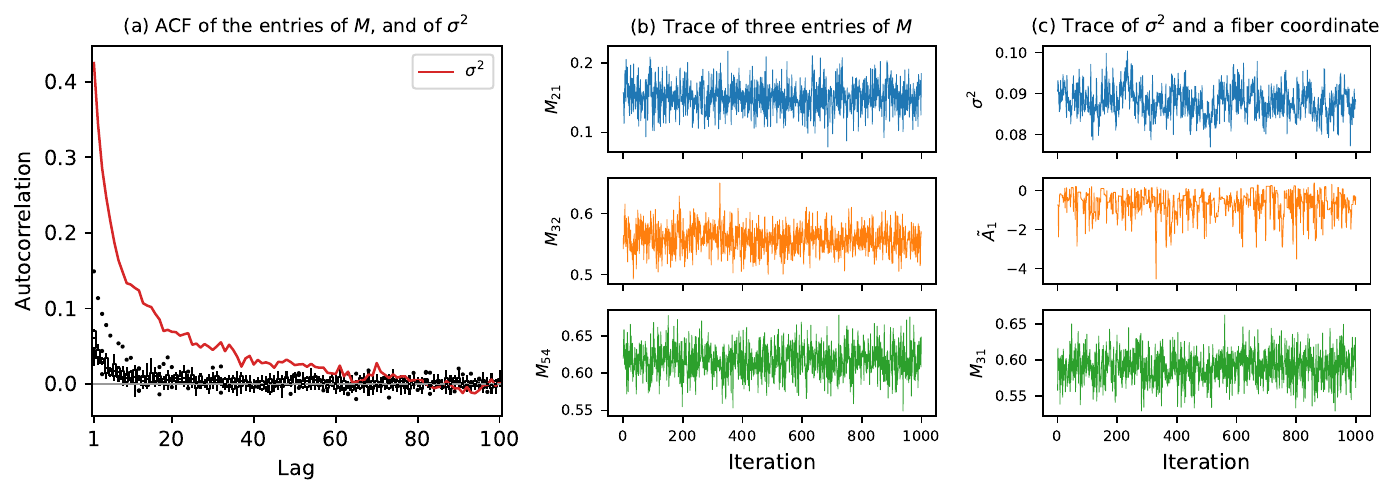}
    \caption{Sampler diagnostics for one chain of $20000$ iterations on a single data set with $n = 200$. (a) Boxplots, over the $\bar d = 10$ entries of $M$, of the autocorrelation at each lag, with the autocorrelation of $\sigma^2$ overlaid. (b) Traces of three entries of $M$ over the first $1000$ iterations. (c) Traces of $\sigma^2$, of a fiber coordinate $\tilde A_i$, and of a further entry of $M$.}
    \label{fig:rank-diagnostics}
\end{figure}

Step (i) updates $A_i$, which is confined to the cone $\{A\preceq0\}$ and is sampled through $A_i = -\exp(\tilde A_i)$ over unconstrained symmetric $\tilde A_i$, as proposed in Example~\ref{ex:nuclear-unit-diag}. Observations are grouped by $d_{0,i}$ so that each group is sampled at its own fixed dimension. We run one chain of $20000$ iterations after $2000$ of burn-in. The adapted step sizes are $1.16$, $0.69$ and $0.49$ for $d_0 = 1,2,3$, with acceptance rates $0.570$, $0.574$ and $0.573$. Figure~\ref{fig:rank-diagnostics} summarizes the output: the effective sample size of the $\bar d$ entries of $M$ is at least $6878$, with median $13583$ out of $20000$ draws, while that of $\sigma^2$ is $1912$.

Table~\ref{tab:sim-rank} reports the recovery of $M$. The scaled error $\sqrt n\,\|\hat M-M_0\|_F$ is flat at about $0.94$ once $n\ge 800$, so the error decays at the $n^{-1/2}$ rate, and $\mathrm E(\sigma^2\mid Y_{1:n})$ approaches the true $0.09$.

\begin{table}[H]
    \centering
    \caption{Rank-varying correlation matrices, one chain of $20000$ iterations per sample size. Errors are computed on the $\bar d = 10$ identified entries of $M$.}
    \label{tab:sim-rank}
    \begin{tabular}{lccc}
        \toprule
        & $n = 200$ & $n = 800$ & $n = 3200$\\
        \midrule
        $\|\hat M-M_0\|_F$                & $0.054$  & $0.033$  & $0.017$\\
        $\sqrt n\,\|\hat M-M_0\|_F$       & $0.757$  & $0.937$  & $0.937$\\
        $\mathrm E(\sigma^2\mid Y_{1:n})$ & $0.1026$ & $0.0887$ & $0.0895$\\
        \bottomrule
    \end{tabular}
\end{table}

\begin{figure}[H]
    \centering
    \includegraphics[width=\textwidth]{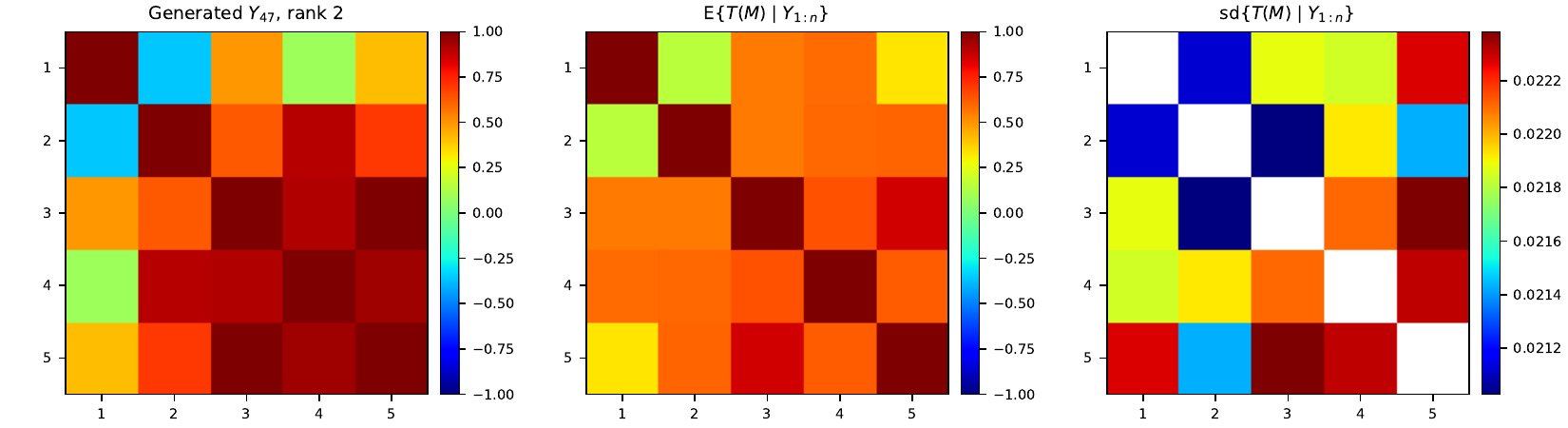}
    \caption{Left: one generated $Y_i$, here of rank $2$. Center and right: the posterior mean and posterior standard deviation of $T(M)$. The left and center panels share a color scale, and entries that are exactly zero are drawn white.}
    \label{fig:mat-rank}
\end{figure}

Figure~\ref{fig:mat-rank} shows one of the rank-deficient observations alongside the posterior mean and standard deviation of $T(M)$, on the same color scale.

\subsubsection*{Sparse correlation matrices}

The last scenario is Example~\ref{ex:sparse-correlation}, the elliptope with an $\ell_1$ penalty on the off-diagonal entries, again with $d = 5$ and a fixed mean. The sparsity pattern is treated as known, so $\lambda_{jk} = \lambda_0 = 0.35$ on three designated entries and $\lambda_{jk} = 0$ elsewhere, and $Z_i$ is generated exactly as in the preceding simulation with $\sigma^2 = 0.09$ and $M$ set to zero on those three entries. The mirror map is now
\(
    T(Z_i) = \arg\min_{W\in\mathcal M}\Bigl\{\tfrac12\|W - Z_i\|_F^2 + \sum\nolimits_{j<k}\lambda_{jk}|W_{jk}|\Bigr\},
\)
which we evaluate by the alternating direction method of multipliers, splitting the smooth and $\ell_1$ terms from the positive semidefinite cone. Of the penalized entries, $41.0\%$ are driven exactly to zero, and the resulting $Y_i$ have ranks $5$, $4$ and $3$ in counts $10$, $159$ and $31$, so the observations vary in sparsity and in rank at once.

\begin{figure}[H]
    \centering
    \includegraphics[width=\textwidth]{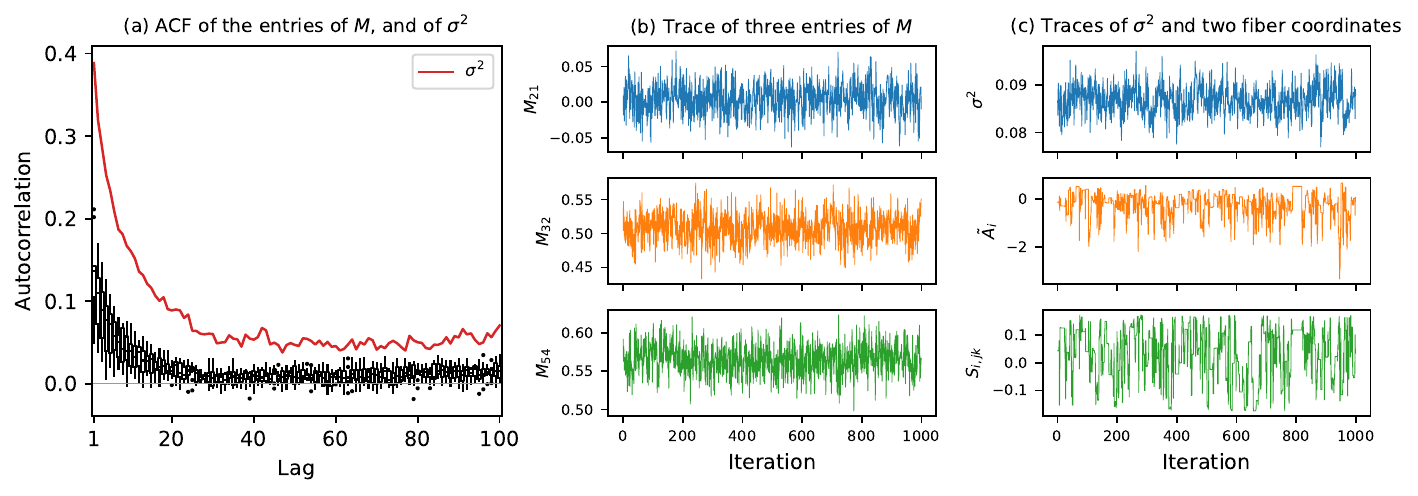}
    \caption{Sampler diagnostics for one chain of $20000$ iterations on a single data set with $n = 200$. (a) Boxplots, over the $\bar d = 10$ entries of $M$, of the autocorrelation at each lag, with the autocorrelation of $\sigma^2$ overlaid. (b) Traces of three entries of $M$ over the first $1000$ iterations. (c) Traces of $\sigma^2$, of a fiber coordinate $\tilde A_i$, and of a box coordinate $S_{i,jk}$ on a zeroed entry.}
    \label{fig:sparse-diagnostics}
\end{figure}

The fiber now has three parts rather than two. Alongside $\nu_i$, which integrates out, and $A_i$ on $\ker Y_i$, the subdifferential $\Gamma(Y_i)$ contributes a coordinate $S_{i,jk}$ on each penalized entry that $T$ has set to zero, confined to the interval $[-\lambda_{jk},\lambda_{jk}]$; on the penalized entries that survive, $S_{i,jk}$ is pinned at $\pm\lambda_{jk}$. We sample the interval coordinates through $S_{i,jk} = \lambda_{jk}\tanh(\tilde s_{i,jk})$ and $A_i$ as in the preceding simulation. Because both $d_{0,i}$ and the number of zeroed entries vary across observations, the data are grouped by that pair, giving twelve groups here. Over one chain of $20000$ iterations after $2000$ of burn-in, the acceptance rate lies between $0.558$ and $0.628$ across the eleven groups that carry a free coordinate. Figure~\ref{fig:sparse-diagnostics} summarizes the output: the effective sample size of the $\bar d = 10$ entries of $M$ is at least $3828$, with median $6527$ out of $20000$ draws, while that of $\sigma^2$ is $654$.

Table~\ref{tab:sim-sparse} reports the recovery of $M$. The scaled error $\sqrt n\,\|\hat M-M_0\|_F$ is flat at about $1.12$ across a sixteen-fold range of $n$, so the error decays at the $n^{-1/2}$ rate, and $\mathrm E(\sigma^2\mid Y_{1:n})$ approaches the true $0.09$.

\begin{table}[H]
    \centering
    \caption{Sparse correlation matrices, one chain of $20000$ iterations per sample size. Errors are computed on the $\bar d = 10$ identified entries of $M$.}
    \label{tab:sim-sparse}
    \begin{tabular}{lccc}
        \toprule
        & $n = 200$ & $n = 800$ & $n = 3200$\\
        \midrule
        $\|\hat M-M_0\|_F$                & $0.081$  & $0.039$  & $0.020$\\
        $\sqrt n\,\|\hat M-M_0\|_F$       & $1.151$  & $1.108$  & $1.124$\\
        $\mathrm E(\sigma^2\mid Y_{1:n})$ & $0.0956$ & $0.0902$ & $0.0901$\\
        \bottomrule
    \end{tabular}
\end{table}

\begin{figure}[H]
    \centering
    \includegraphics[width=\textwidth]{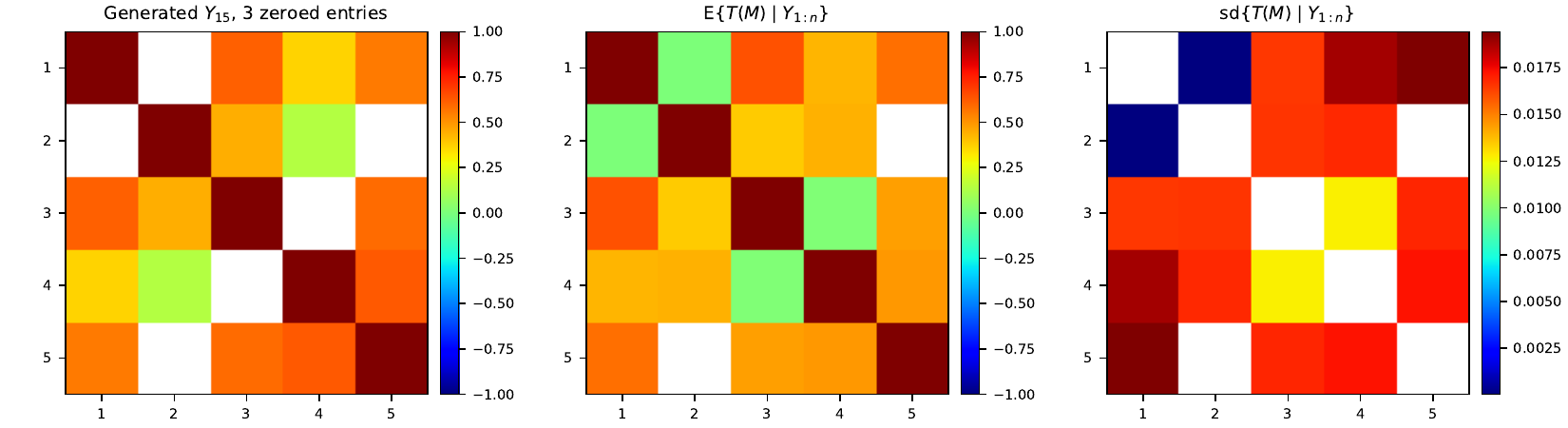}
    \caption{Left: one generated $Y_i$, here with all three penalized entries set exactly to zero and drawn white. Center and right: the posterior mean and posterior standard deviation of $T(M)$, on the same color scale as the left panel. Entries that are exactly zero, including those of $T(M)$ that vanish in every draw, are drawn white.}
    \label{fig:mat-sparse}
\end{figure}

Figure~\ref{fig:mat-sparse} shows one observation whose three penalized entries were all set exactly to zero, alongside the posterior mean and standard deviation of $T(M)$. An entry that $T(M)$ zeroes in every draw has posterior standard deviation exactly zero, and appears white in both of the right-hand panels.

\end{document}